\documentclass[11pt]{article}
\usepackage{amsmath,amsfonts,amsthm,amssymb}

\newcommand\version{Dec. 22, 2004}

\numberwithin{equation}{section}

\newcommand\N{{\mathbb N}}
\newcommand\C{{\mathbb C}}
\newcommand\SsFF{{\widetilde S}}
\newcommand\R{{\mathbb R}}
\newcommand\eps{\varepsilon}
\newcommand\half{\mbox{$\frac 12$}}
\newcommand\rhov\varrho

\newcommand\const{{\rm const.\, }}
\newcommand\Tr{{\rm Tr}}
\newcommand\Ff{{\cal F}}
\newcommand\Ecal{{\cal E}}
\newcommand\Hh{{\cal H}}
\newcommand\hN{{\hat N}}
\newcommand\Pp{{\cal P}}
\newcommand\gamo{\gamma}
\newcommand\upa{\uparrow}
\newcommand\doa{\downarrow}
\newcommand\per{{\rm per}}
\newcommand\vu{\nu}

\renewcommand\kappa\varkappa
\renewcommand\rho\varrho
\newcommand\AaFF{{\mathbb A}}
\newcommand\id{{\mathbb I}}

\def\Xint#1{\mathchoice 
{\XXint\displaystyle\textstyle{#1}}%
{\XXint\textstyle\scriptstyle{\raise 2pt\hbox{\tiny{$#1$}}}}%
{\XXint\scriptstyle\scriptscriptstyle{#1}}%
{\XXint\scriptscriptstyle\scriptscriptstyle{#1}}%
\!\int} 
\def\XXint#1#2#3{{\setbox0=\hbox{$#1{#2#3}{\int}$} 
\vcenter{\hbox{$#2#3$}}\kern-.5\wd0}} 
\def\intsum{\Xint\sum}

\newtheorem{thm}{THEOREM}
\newtheorem{lem}{Lemma}
\newtheorem{cor}{COROLLARY}

\begin{document}

\markboth{\scriptsize{RS \version}}{\scriptsize{RS \version}}
\title{\bf{THE THERMODYNAMIC PRESSURE\\ OF A DILUTE FERMI GAS}}
\author{\vspace{5pt} Robert Seiringer
\\ \vspace{-2pt}\small{ Department of Physics, Jadwin Hall, Princeton
University,  }\\ \vspace{-2pt}\small P.O. Box 708, Princeton NJ
08544, USA. 
\\ {\small Email: \texttt  {rseiring@math.princeton.edu}} }
\date{\small \version}
\maketitle

\begin{abstract}
  We consider a gas of fermions with non-zero spin at temperature $T$
  and chemical potential $\mu$. We show that if the range of the
  interparticle interaction is small compared to the mean particle
  distance, the thermodynamic pressure differs to leading order from
  the corresponding expression for non-interacting particles by a term
  proportional to the scattering length of the interparticle
  interaction. This is true for any repulsive interaction, including
  hard cores. The result is uniform in the temperature as long as $T$
  is of the same order as the Fermi temperature, or smaller.
\end{abstract}

\renewcommand{\thefootnote}{${\,}$}
\footnotetext{Work partially supported by U.S. National Science
Foundation grant PHY-0353181 and by an Alfred P. Sloan Fellowship.}
\renewcommand{\thefootnote}{${\, }$}
\footnotetext{\copyright\,2004 by the author.
This paper may be reproduced, in its entirety, for non-commercial
purposes.}


\section{Introduction and Main Results}

The physics of dilute gases at low temperature has received a lot of
interest in the last couple of years, due to the recent experimental
advances in studying these systems. Despite tremendous interest in the
problem, rigorous results starting from first principles remain
sparse, and often one has to rely on uncontrolled approximations to
obtain quantitative information. This is true especially for {\it
  dilute} systems, where the interparticle interaction can not easily
be taken into account using perturbation theory. Here, dilute refers
to the case when the range of the interparticle interaction is small
compared with the mean particle distance.

The first, although admittedly not the most interesting, question to
ask is for the ground state energy of the system. In \cite{LY98},
Lieb and Yngvason devised a method for proving the relevant expression
for dilute Bose gases. In this case, the energy per unit volume at
density $\rho$ is given by $ 4\pi a \rho^2$, where $a$ is the ($s$-wave) 
scattering length of the interaction potential, and $a^3\rho\ll
1$, i.e., the system is dilute. (Units are chosen such that $\hbar=1$
and $2m=1$, where $m$ denotes the mass of the particles.)  The
corresponding expression for a two-dimensional Bose gas was later
proved in \cite{LY01}.

Recently, it was possible to extend these methods and prove the
corresponding result for fermions \cite{LSS}. That is, the ground
state energy density of a dilute gas of spin $q$ fermions is given by
\begin{equation}\label{eqferm}
\frac 35 \left(\frac {6\pi^2}q\right)^{2/3} \rho^{5/3} 
+ 4\pi a \rho^2 \left(1-q^{-1}\right) +\text{ higher order in } (a^3\rho) \,.
\end{equation}
As before, $a$ denotes the scattering length. The factor $(1-q^{-1})$ in the
interaction energy results from the fact that only particles with
different spin can exhibit $s$-wave scattering. The contribution from
the interaction between particles of the same spin is of higher order
in $\rho$.

In this paper, we prove the analogue of (\ref{eqferm}) at positive
temperature. Our main result is Theorem~\ref{T1}. We work in the grand
canonical ensemble, and consider the pressure of the gas at given
temperature $T$ and chemical potential $\mu$. We will show that, for
dilute gases, the effect of the particle interaction results in  a
contribution $-4\pi a \rho^2 (1-q^{-1})$ to the pressure, where $\rho$
is now the average density. This result holds for any temperature, as
long as $T$ is not much bigger than the Fermi temperature (for the
non-interacting gas), given by $T_{\rm F}= (6\pi^2/q)^{2/3}
\rho^{2/3}$ (in units where $k_{\rm B}=1$). The rational behind this
formula is the following: for dilute gases, the effect of the
interaction reduces to two-particle $s$-wave scattering, which can
take place only between particles of unequal spin. This is just like
in the zero-temperature (ground state) case. The effect of the
temperature on this scattering process is negligible, since for
$T\lesssim T_{\rm F}$, the thermal wave length is of the same order
(or greater) than the mean-particle distance. The aim of this
paper is to make this intuition precise.

\bigskip
We will now describe the system in detail. For simplicity, we consider
here only the case $q=2$, i.e., the spin $1/2$ case. The extension to
$q>2$ is straightforward.  The Hilbert space under consideration is
given by the fermionic Fock space for spin $1/2$ particles, $\Ff=
\Ff_{\rm F}(L^2(\Lambda_L;\C^2))$. Here, $\Lambda_L$ denotes a cube of
side length $L$. The Hamiltonian is the direct sum
$H=\bigoplus_{N=0}^\infty H_N$, with $H_0=0$, $H_1=- \Delta$, and
\begin{equation}
H_N=\sum_{i=1}^N -\Delta_i + \sum_{1\leq i<j\leq N} v(x_i-x_j)
\end{equation}
for $N\geq 2$.  Here, $\Delta$ denotes the Laplacian with Dirichlet
boundary conditions on $\Lambda_L$. Units are chosen such that $\hbar=1$
and $2m=1$, where $m$ denotes the mass of the particles. We note that
both $H$ and $\Ff$ depend on $L$, of course, but we suppress this
dependence in our notation.

The pair potential $v$ is assumed to be positive, radial, and of
finite range $R_0$. It then has a finite and positive scattering
length $a$. The scattering length may be defined as follows: if
$\varphi$ is the unique solution of the zero-energy scattering
equation
\begin{equation}\label{scatteq}
-\Delta \varphi + \half v \varphi = 0
\end{equation}
subject to the boundary condition $\lim_{|x|\to\infty} \varphi(x)=1$,
then $a$ is given by $a= \lim_{|x|\to\infty} |x| (1- \varphi(x))$ (see
Appendix~A in \cite{LY01} for details).  Note that we do not assume $v$ to be
integrable, our results also apply to the case of a hard core. Note
also that for a pure hard-core interaction, the scattering length is
equal to the range.

Our main result concerns the pressure of the system at some given
inverse temperature $\beta=1/(k_{\rm B} T)$ and chemical potential $\mu$. It is
given by
\begin{equation}\label{pre}
P(\beta,\mu) = \lim_{L\to\infty} \frac 1{L^3 \beta} \ln
\Tr_\Ff\, \exp\big(-\beta(H-\mu \hN)\big)\,.
\end{equation}
Here, $\hN$ denotes the number operator. It is well known that for
systems with short range interactions the limit in (\ref{pre}) exists
and is independent of boundary conditions \cite{robinson,ruelle}. 

We are interested in $P(\beta,\mu)$ for low density, $\rhov$, which is
given by $\rhov=\partial P/\partial\mu$, assuming the derivative
exists. (Note that $P(\beta,\mu)$ is convex in $\mu$ and therefore the
derivative exists almost everywhere. In particular, the right and left
derivatives exist.)  With {\it low density} we mean that the
dimensionless quantity $a^3\rhov$ is small, i.e., that the gas is {\it
  dilute}. Note that there are two dimensionless quantities in this
problem: $a^3\rhov$, measuring the diluteness, and the fugacity
$z=e^{\beta \mu}$. Small $z$ corresponds to the (high temperature)
limit of a classical gas, whereas for large $z$ the system approaches
its ground state. Since we are interested in the quantum aspects of
the system, we will consider the case when $1/z$ is bounded.  Another
way of saying the same thing is that there are three length scales in the
problem: the scattering length $a$, the mean particle distance
$\rhov^{-1/3}$, and the thermal wavelength $\beta^{1/2}$. We are
interested in the regime where $a \ll \rhov^{-1/3} \lesssim
\beta^{1/2}$, i.e., where $a$ is much smaller than $\rhov^{-1/3}$ and
$\rhov^{-1/3}$ is comparable with, or much smaller than,
$\beta^{1/2}$.

Let $P_0(\beta,\mu)$ and $\rhov_0(\beta,\mu)$ be the pressure and
density of a non-interacting gas of spin $1/2$ fermions. They
are given by
\begin{equation}\label{pzero}
P_0(\beta,\mu)=\frac{2}{\beta} (2\pi)^{-3} \int_{\R^3} dp\,
\ln\left(1+z\exp(-\beta p^2)\right)
\end{equation}
and
\begin{equation}\label{rhozero}
\rhov_0(\beta,\mu)= \frac{\partial P_0(\beta,\mu)}{\partial\mu}= 
2 (2\pi)^{-3} \int_{\R^3} dp\, \frac 1{1+z^{-1}
\exp(\beta p^2)},
\end{equation}
respectively. The factor $2$ in front of the integrals takes the
number of spin states into account.

Our main result is the following.

\begin{thm}\label{T1}
  Let $\rhov_0\equiv \rhov_0(\beta,\mu)$ and let $z=e^{\beta \mu}$.
  For any $\alpha< 1/33$ there exists a function $C_\alpha(z)$,
  uniformly bounded in $z$ for bounded $1/z$, such that
\begin{equation}\label{thm1}
\left| P(\beta,\mu) - P_0(\beta,\mu) + 2\pi a \rhov_0(\beta,\mu)^2
\right| \leq C_\alpha(z) a\rhov_0^2 \big( a\rhov_0^{1/3}\big)^{\alpha}\,.
\end{equation}
\end{thm}

We remark that the non-uniformity of our bound for small $z$ is not
an artifact of our method of proof. In the limit $z\to 0$ one obtains
a classical gas where $2\pi a \rhov_0^2$ certainly does {\it not} give
the correct contribution of the interaction at low density.  In particular, this
term depends on Planck's constant (which equals 1 in our units).

Our proof below gives explicit bounds on the $C_\alpha(z)$ appearing
in the statement of the theorem. Since the exponent $\alpha$ in the
error term on the right side of (\ref{thm1}) is far from optimal,
however, we did not make the effort of writing down these bounds
explicitly. Moreover, $C_\alpha(z)$ depends on the interaction
potential $v$ only through its range or, more precisely, through the
dimensionless ratio $R_0/a$.  Also this dependence could in principle
be given explicitly.

We note that $\rhov_0(\beta,\mu)$ can be replaced by the true density
(of the interacting gas) in Eq.~(\ref{thm1}). In fact, we have the
following corollary of Theorem~\ref{T1}.

\begin{cor}\label{C1}
  Let $\rhov_\pm(\beta,\mu) = \partial P(\beta,\mu)/\partial\mu^\pm$
  denote the right and left derivatives of $P(\beta,\mu)$,
  respectively.  For any $\alpha< 1/33$ there exists a function
  $\widehat C_\alpha(z)$, uniformly bounded in $z$ for bounded $1/z$, such that
\begin{equation}\label{thm2}
\frac{\left|\rhov_\pm(\beta,\mu) -
\rhov_0(\beta,\mu)\right|}{\rhov_0(\beta,\mu)}\leq
\widehat C_\alpha(z) \big( a\rhov_0^{1/3}\big)^{(1+\alpha)/2}\,.
\end{equation}
\end{cor}

The proof of Theorem~\ref{T1} will actually not only show that the
density of the interacting and non-interacting system are close, as
claimed in (\ref{thm2}), but also the reduced one-particle density
matrices (cf. Eq.~(\ref{summar})).  It is to expect that
$\rhov_\pm(\beta,\mu)=\rhov_0(\beta,\mu)-2\pi a
\partial\rhov_0(\beta,\mu)/\partial \mu \,+ $ higher order terms in
$a^3\rhov_0$, but we do not have a proof of this claim. Nevertheless,
we note that Theorem~\ref{T1} implies a similar statement for other
thermodynamic potentials. We state the following corollary for the
Helmholtz free energy, but an analogous result holds for other
thermodynamic potentials as well.

\begin{cor}\label{C2}
  For $\rho>0$, let $f(\beta,\rho) = \sup_{\mu}[ \mu\rho - P(\beta,\mu)]$ denote
  the free energy density, and $f_0(\beta,\rho)$ the
  corresponding quantity for the non-interacting system. For any
  $\alpha< 1/33$ there exists a function $\widetilde C_\alpha(x)$, uniformly
  bounded in $x$ for bounded $1/x$, such that
\begin{equation}\label{thm3}
\left| f(\beta,\rho) - f_0(\beta,\rho) - 2\pi a \rho^2
\right| \leq \widetilde C_\alpha\big(\beta\rho^{2/3}\big) a\rho^2 
\big( a\rho^{1/3}\big)^{\alpha}\,.
\end{equation}
\end{cor}

\bigskip

Theorem~\ref{T1} (and Corollaries~\ref{C1} and~\ref{C2}) can be
extended in several ways, as will be explained now. For simplicity, we omit
the proof of these assertions here. They can be proved by only
small modifications of the proof of Theorem~\ref{T1} given below,
which is already quite lengthy itself.

\begin{itemize}
\item {\it Polarized gas:} A term $m S^3_{\rm tot}$ can be added to
  the Hamiltonian, where $S^3_{\rm tot}$ denotes the 3-component of
  the total spin of the particles, and $m$ is a coupling parameter
  proportional to the magnetic field. This has the effect of adding a
  \lq spin-dependent\rq\ chemical potential. Theorem~\ref{T1} also holds
  in this case (with $P$ and $P_0$ depending on $m$, of course), if
  $2\pi a \rhov^2$ is replaced by $8\pi a \rhov_\doa \rhov_\upa$. Here,
  $\rhov_\doa$ and $\rhov_\upa$ denote the density of spin-up and
  spin-down particles, respectively. They are given by derivatives of
  the pressure as $\half \partial P/\partial\mu \pm  \partial P/\partial
  m$.
  
\item {\it Higher spin:} The case of higher spin can be treated in the
  same way. If $q$ denotes the number of spin states, then the leading
  order contribution of the interaction energy per unit volume for a
  dilute gas is $4\pi a (1-q^{-1})\rhov^2$, which reduces to
  (\ref{thm1}) in the case $q=2$. Also the polarized gas can be
  studied for spin higher then $1/2$.
  
\item {\it Infinite range potentials:} As already mentioned, the error
  term on the right side of (\ref{thm1}) depends on the interaction
  potential only through the ratio of its range to its scattering
  length, $R_0/a$. By cutting off the potential $v$ in an appropriate
  ($\rhov$-dependent) way, it is therefore possible to extend
  Theorem~\ref{T1} to infinite range potentials (with finite
  scattering length), with possibly a worse error term than the one
  given in (\ref{thm1}).  (Compare with the corresponding discussion
  for the Bose gas in the appendices of \cite{LSY00} and
  \cite{LY01}.)
  
\item {\it The two-dimensional gas:} A corresponding result can also
  be derived for a Fermi gas in two dimensions. The leading
  contribution of the interaction energy for a dilute gas is then
  $2\pi \rhov^2 /|\ln a^2\rhov|$ per unit volume. This was shown in
  \cite{LSS} for the ground state, i.e., at zero temperature, and the
  methods developed in this paper can be used to obtain this result
  also at positive temperature.

\end{itemize}

Before giving the full proof of Theorem~\ref{T1}, we start with a
short outline to guide the reader. In the next Section~\ref{prelim},
we state some preliminaries that will be useful for our proofs. We
introduce the pressure functional which defines a variational
principle for the pressure. We also state some useful properties of
the non-interacting system. Sections~\ref{LB} and~\ref{UB} contain the
proof of Theorem~\ref{T1}. This proof is divided into two parts, the
lower and upper bounds to the pressure. Finally, in Section~\ref{CP},
we give the proof of Corollaries~\ref{C1} and~\ref{C2}.

For the lower bound to the pressure, given in Section~\ref{LB}, it is
necessary to construct an appropriate trial density matrix for the
pressure functional. As in the zero temperature calculation in
\cite{LSS}, we find it necessary to choose a trial density matrix that
confines particles into small boxes in order to control the average
particle number in each box. The construction in each small box is
done in Subsect.~\ref{ssconst}. We then proceed with the calculation
of the variational pressure in Subsect.~\ref{sscalc}. We use similar
methods as in \cite{LSS} to estimate the energy of the state. In
addition, it is necessary to estimate its entropy, which is done with
the aid of Lemma~\ref{Lem1}.  The price one has to pay for using the
box method are finite size corrections, which are estimated in
Subsect.~\ref{sstrace}. The final result is then stated in
Subsect.~\ref{ssfinal}.

The upper bound to the pressure, given in Section~\ref{UB}, has two
main ingredients. First, an  operator inequality proved in
\cite[Lemma~4]{LSS}, which allows for the replacement of the
interaction potential $v$ by a \lq\lq soft\rq\rq\ potential $U$, at
the expense of the high-momentum part of the kinetic energy. See
Subsect.~\ref{sslow}. Secondly, an improved version of subadditivity
of entropy in Subsect.~\ref{ssl2}, which was recently derived in
\cite{LSent}. As shown in Subsect.~\ref{apse}, this estimate allows to
prove that the reduced one-particle density matrix of the spin-up
particles, for {\it fixed} positions of the spin-down particles, is
close to the Fermi-Dirac distribution for non-interacting particles.
This property of the reduced one-particle density matrix is then used
to show, in Subsect.~\ref{ssl4}, that first order perturbation theory
with the soft potential $U$ gives the correct answer for dilute gases.

\section{Preliminaries}\label{prelim}

Since the Hamiltonian $H$ does not depend on the spin variables and,
in particular, commutes with the operators counting the number of
spin-up and spin-down particles, the problem can be reformulated in
terms of two species of spinless fermions. More precisely, $\Ff\cong
\Ff_{\rm F}(\Hh_1)\otimes\Ff_{\rm F}(\Hh_1)\equiv \Ff_\upa\otimes
\Ff_\doa$, where $\Hh_1=L^2(\Lambda_L;\C)$ denotes the one-particle
space for {\it spinless} particles. We label particle coordinates in
the first factor by $x^\upa$ and in the second by $x^\doa$. The
Hamiltonian in this representation can then be written as
$H=\bigoplus_{N,M=0}^\infty H_{N,M}$, with
\begin{eqnarray}\nonumber 
H_{N,M} &=& - \sum_{i=1}^N
\Delta^\upa_{i}- \sum_{k=1}^M \Delta^\doa_{k} + \sum_{i=1}^N
\sum_{k=1}^M v(x^\upa_i-x^\doa_k) \\ && + \sum_{1\leq i<j\leq N}
v(x^\upa_i-x^\upa_j)+ \sum_{1\leq k<l\leq M} v(x^\doa_k-x^\doa_l)
\,.\label{ham2} 
\end{eqnarray} 
The first two terms are simply the kinetic energies of the spin-up and
spin-down particles, and the interaction potential is divided into
three parts, corresponding to interaction between particles of the
same and of different spin, respectively.  In a sector of fixed
particle numbers $N$ and $M$, we denote the particle coordinates
collectively by $X^\upa=(x^\upa_1,\dots,x^\upa_N)$ and
$X^\doa=(x^\doa_1,\dots,x^\doa_M)$.

\subsection{The Pressure Functional}

The pressure (\ref{pre}) can be computed via a variational
principle. For $\Gamma$ a density matrix, i.e., a positive trace class
operator on $\Ff$ with $\Tr_\Ff\, \Gamma=1$, we define the pressure
functional $\Pp^L[\Gamma]$ by
\begin{equation}\label{func}
-L^3\, \Pp^L[\Gamma]= \Tr_\Ff  \left[ ( H-\mu \hN)\Gamma\right] -
\frac 1\beta S[\Gamma]\,,
\end{equation}
where $S[\Gamma]= - \Tr_\Ff\, (\Gamma \ln \Gamma)$ denotes the (von
Neumann) entropy. (The expression (\ref{func}) is well defined if the
eigenfunctions of $\Gamma$ are in the quadratic form domain of $H$ and
$\hN$. Otherwise, we can take it to be $+\infty$.)  Let
$P^L(\beta,\mu)$ denote the maximum of $\Pp^L[\Gamma]$ over all
density matrices. The maximum is uniquely attained by the
grand-canonical Gibbs density matrix, given by $\exp(-\beta(H-\mu
\hN))/\Tr_\Ff \exp(-\beta(H-\mu \hN))$. Hence the pressure
$P^L(\beta,\mu)$ is given by $P^L(\beta,\mu)= L^{-3}\beta^{-1}\ln
\Tr_\Ff\, \exp\left(-\beta(H-\mu \hN)\right)$, and
$P(\beta,\mu)=\lim_{L\to\infty} P^L(\beta,\mu)$.

At zero temperature, i.e., when $\beta=\infty$, this variational principle
reduces to the usual variational principle for the ground state
energy. Note, however, that at positive temperature the functional
(\ref{func}) is not linear in the density matrix.

\subsection{The Ideal Fermi Gas}

For later use, we also define the pressure functional for the
noninteracting gas, $\Pp_0^L[\Gamma]$. It is defined in the same way
as $\Pp^L[\Gamma]$ above, with $H$ replaced by the non-interacting
Hamiltonian $H^{(0)}=\bigoplus_{N,M=0}^\infty H^{(0)}_{N,M}$, where
$H^{(0)}_{N,M}$ is given as in (\ref{ham2}) but with $v=0$. We denote
the (finite volume) pressure for the non-interacting gas by
$P_0^L(\beta,\mu)$. It is given by 
\begin{equation}\label{pzerofin}
P_0^L(\beta,\mu)=\frac{2}{\beta} \frac 1{L^3} \,\Tr_{\Hh_1}\,
\ln\big(1+z\exp(\beta \Delta)\big)\, ,
\end{equation}
which reduces to (\ref{pzero}) in the thermodynamic limit. We note that this
expression can also be obtained from a variational principle for the
reduced one-particle density matrix (see, e.g., \cite{Thirring}).
Namely,
\begin{equation}\label{onepf}
-\half L^3 P^L_0(\beta,\mu) = \inf_{\gamma} \left[ \Tr_{\Hh_1} 
(-\Delta-\mu)\gamma - \frac 1\beta \SsFF[\gamma] \right]\,,
\end{equation}
where the infimum is over all positive trace class operators $\gamma$ on $\Hh_1$ 
with $0\leq \gamma\leq 1$, and $\SsFF[\gamma]$ is given by
\begin{equation}\label{defss}
\SsFF[\gamma]= \Tr_{\Hh_1}\left[ - \gamma\ln\gamma - (1-\gamma)\ln(1-\gamma)\right]\,.
\end{equation}  
The reason for the factor $1/2$ in front of $P_0^L$ are the $2$
different spin states, which we have not accounted for in the
functional. The infimum in (\ref{onepf}) is uniquely attained at
$\gamma_0=(1+z^{-1} e^{-\beta\Delta})^{-1}$.

\section{Lower Bound to the Pressure}\label{LB}

We start the proof of Theorem~\ref{T1} by deriving a lower bound to
the pressure. Since $P(\beta,\mu)$ is determined by
maximizing the pressure functional, a lower bound can be derived using
an appropriate trial density matrix in the pressure functional
(\ref{func}).

\subsection{The Box Method}

It will be convenient to divide space into small boxes of side length
$\ell$ and confine the particles to these boxes. By choosing $\ell$
appropriately, we can then control the average particle number in
every box. Moreover, if we keep these boxes separated by a distance
$R_0$, there is no interaction between particles in different boxes.

More precisely, pick an integer $I$ and divide the interval $[0,L]$
into $I$ intervals of equal length. We choose $I$ such that $\ell
\equiv L/I - R_0>0$. From the variational principle defined by
(\ref{func}) we can infer that
\begin{equation}\label{boxes}
L^3 P^L(\beta,\mu) \geq I^3 \ell^3 P^\ell(\beta,\mu)\,,
\end{equation}
where the factor $I^3$ is the number of boxes. Dividing (\ref{boxes})
by $L^3$ and letting $L\to \infty$ and $I\to \infty$ in such a way
that $L/I$ converges to some number greater than $R_0$, we see that 
\begin{equation}\label{boxbound}
P(\beta,\mu) \geq \frac 1{(1+R_0/\ell)^3} P^\ell(\beta,\mu)
\end{equation}
for any $\ell>0$. 

\subsection{Construction of the Trial Density Matrix}\label{ssconst}

We now construct a trial density matrix for $\Pp^\ell[\Gamma]$. For
fixed $\beta$ and $\mu$, let $\rhov_0=\rhov_0(\beta,\mu)$, and let
$K>0$. Let $Q$ be the projector onto the subspace of
$\Hh_1=L^2(\Lambda_\ell;\C)$ where $-\Delta \leq K\rhov_0^{2/3}$,
i.e., $Q=\theta(K\rhov_0^{2/3}+\Delta$). Here, $\theta$ denotes the
Heaviside step function, given by
\begin{equation}
\theta( t) = \left\{ \begin{array}{ll} 0 & {\rm for\ } t<0 \\ 
1 & {\rm for\ } t\geq 0 \, . \end{array} \right.
\end{equation}
On $\Ff_Q\equiv \Ff(Q \Hh_1)\otimes \Ff(Q \Hh_1)$, let $H_Q$ denote
the second quantization of $-\Delta Q$, and let $\Gamma_Q$ be the
corresponding grand canonical Gibbs density matrix, defined as
$\Gamma_Q= \exp(-\beta(H_Q-\mu \hN))/\Tr_{\Ff_Q}\exp(-\beta(H_Q-\mu
\hN))$. We use the same symbol for the density matrix on $\Ff$, being
$\Gamma_Q$ on the subspace $\Ff_Q$ and $0$ on the orthogonal
complement. Let $\rhov_Q=\ell^{-3}\Tr_\Ff \hat N \Gamma_Q$ be the
average density of $\Gamma_Q$. By explicit computation,
\begin{equation}
\rhov_Q= \frac 2{\ell^3} \Tr_{\Hh_1}  \, Q \frac 1{1+\exp(\beta (
-\Delta - \mu))}\,.
\end{equation}

We can decompose $\Gamma_Q$ as
\begin{equation}
\Gamma_Q = \sum_{\alpha} \lambda_\alpha |\psi_\alpha\rangle\langle
\psi_\alpha|\,,
\end{equation}
with $\lambda_\alpha\geq 0$, $\sum_\alpha \lambda_\alpha =1$, and
$\{\psi_\alpha\}$ an orthonormal set in $\Ff_Q$. Moreover, we can
always choose the $\psi_\alpha$ to be products of Slater determinants
of $N_\alpha$ $\upa$-particles and $M_\alpha$ $\doa$-particles,
respectively, for some $N_\alpha,M_\alpha\in\N$.  Given such a
$\psi_\alpha$, we define
\begin{equation}\label{defp1}
\phi_\alpha = \frac {F^{N_\alpha,M_\alpha}
\psi_\alpha}{\|F^{N_\alpha,M_\alpha}\psi_\alpha\|}\,,
\end{equation}
with $F^{N,M}$ given as follows. Pick some $s> 2R_0$ and let
$g:\R^3\mapsto \R$ be function with $0\leq g\leq 1$, having the
property that $g(x)=0$ for $|x|\leq s$ and $g(x)=1$ for $|x|\geq 2s$.
We may also assume that $|\nabla g|\leq \const s^{-1}$ for some
constant independent of $s$. Moreover, for some $\half s \geq R >
R_0$, let $f:\R^3\mapsto \R$ be given by $f(x)=\varphi(x)/(1-a/R)$ for
$|x|\leq R$ and $1$ otherwise. Here, $\varphi$ denotes the solution to
the zero-energy scattering equation (\ref{scatteq}).  Note that $f$ is
a continuous function, since $\varphi(x)= 1 - a/|x|$ for $|x|\geq
R_0$. We define
\begin{equation}\label{defF}
F^{N,M}(X^\upa,X^\doa)=\!\! \prod_{1\leq i<j\leq N}
g(x^\upa_i-x^\upa_j) \prod_{1\leq k<l\leq M}
g(x^\doa_k-x^\doa_l) \prod_{i=1}^{N} \prod_{k=1}^{M}
f(x^\upa_i-x^\doa_k)\,.
\end{equation}

As a trial density matrix for $\Pp^\ell[\Gamma]$ we choose
\begin{equation}\label{defgamma}
\Gamma = \sum_{\alpha} \lambda_\alpha |\phi_\alpha\rangle\langle
\phi_\alpha|\,,
\end{equation}
with $\phi_\alpha$ defined by (\ref{defp1}) and (\ref{defF}). 
Note that the $\phi_\alpha$ are not orthogonal, but they are
normalized, and hence $\Tr_\Ff\, \Gamma = 1$.

\subsection{Calculation of the Variational Pressure}\label{sscalc}

We now derive a lower bound on the variational pressure
$\Pp^\ell[\Gamma]$, with $\Gamma$ given in (\ref{defgamma}). We start
with the expectation value of the energy,
\begin{equation}\label{ener}
\Tr_\Ff\, H \Gamma = \sum_\alpha \lambda_\alpha \langle \phi_\alpha | 
H_{N_\alpha,M_{\alpha}} |\phi_\alpha\rangle\,,
\end{equation}
which we have to bound from above.  This expression can be estimated
using the same methods as in \cite[Sect.~IV]{LSS}.  More precisely, the
calculation in \cite{LSS} shows the following.

\begin{lem}\label{enlem}
  For $N,M\geq 0$, let $D_1$ and $D_2$ denote Slater determinants of
  $N$ and $M$ orthonormal eigenfunctions of the Dirichlet Laplacian on
  a cube of side length $\ell$, respectively. Let $k$ denote the
  maximal kinetic energy of these $N+M$ functions. Let
  $\psi(X^\upa,X^\doa)=D_1(X^\upa) D_2(X^\doa)
  F^{N,M}(X^\upa,X^\doa)$, with $F^{N,M}$ given in (\ref{defF}). Then,
  for some constant $c>0$,
\begin{eqnarray}\nonumber
\frac{\langle \psi| H_{N,M} | \psi\rangle}{\langle\psi|\psi\rangle} 
\!\!\!\! & \leq & \!\!\!\!  \langle D_1 D_2| H^{(0)}_{N,M} | D_1 D_2\rangle + 
8\pi a \frac {N M}{\ell^3} \left( 1 + c N^{-1/3} + c M^{-1/3} \right) 
\\ \nonumber && \!\!\!\!  + c a \frac {N M}{\ell^3}  E\big(R,s,N+M,k,\ell\big) 
+ c (N+M)^{7/3}\frac{s^{3/2} a^{1/2}}{\ell^4} \,, \\ 
\end{eqnarray}
where $E$ is the function
\begin{equation}
E(R,s,n,k,\ell)= \frac {a R^2}{s^3}+s^2 k+\frac aR +  n^{8/3} (s/\ell)^{5}\,.
\end{equation}
\end{lem}

This lemma was proved in \cite{LSS} for the special case when $D_1$
and $D_2$ are Slater determinants of the lowest $N$ and $M$
eigenfunctions of the Dirichlet Laplacian, respectively (compare with
\cite[Eq.~(46)]{LSS}). The proof, however, goes through without change
in the general case, the only difference being the maximal value of
the kinetic energy, $k$, which enters the bound through the estimate
in \cite[Lemma~2]{LSS}. More precisely, the value $n^{2/3}/\ell^2$ in
\cite[Lemma~2]{LSS} has to be replaced by $k$ in the general case
considered here.

Note that the error term in Lemma~\ref{enlem} is not uniform in the
particle number. For this reason, it is necessary to confine the
particles into small boxes, as done here.

In the case of interest in (\ref{ener}), the number of particles is at
most $N+M\leq 2\, \Tr_{\Hh_1} Q \leq c \ell^3 K^{3/2}\rhov_0$ for some
constant $c>0$, and the kinetic energy of each factor in the Slater
determinants is at most $k\leq K \rhov_0^{2/3}$. Hence
Lemma~\ref{enlem} implies the upper bound
\begin{eqnarray}\nonumber
\langle\phi_\alpha| H_{N_\alpha,M_\alpha} |\phi_\alpha\rangle
\!\!\!\! &\leq& \!\!\!\!  \langle\psi_\alpha|H_{N_\alpha,M_\alpha}^{(0)}|\psi_\alpha\rangle
 + 8\pi a\frac {N_\alpha M_\alpha}{\ell^3}\left( 1 +
c N_\alpha^{-1/3}+c M_\alpha^{-1/3} \right) \\ \nonumber && \!\!\!\! 
+ c a \frac {N_\alpha M_\alpha}{\ell^3}
E\big(R,s,K^{3/2}\rhov_0 \ell^3,K\rhov_0^{2/3},\ell \big) \\  && \!\!\!\! 
+ c(N_\alpha+M_\alpha) K^2 \rhov_0^{4/3}s^{3/2}a^{1/2} 
\end{eqnarray}
for some constant $c>0$. We now insert this estimate into (\ref{ener}). We have
\begin{equation}
\sum_\alpha \lambda_\alpha N_\alpha M_\alpha = \Tr_\Ff\, \hN^\upa
\hN^\doa \Gamma_Q = \left(\half {\ell^3}\rhov_Q\right)^2\,,
\end{equation}
where $\hN^\updownarrow$ denotes the number operator on
$\Ff_\updownarrow$. Moreover, using convexity of $x\mapsto x^{3/2}$,
it follows from Jensen's inequality that
\begin{equation}
\sum_\alpha \lambda_\alpha N_\alpha^{2/3} M_\alpha \leq
\left(\half {\ell^3}\rhov_Q\right)^{5/3}\,,
\end{equation}
and likewise with $N_\alpha$ and $M_\alpha$ interchanged.  Also
$\sum_\alpha \lambda_\alpha(N_\alpha+M_\alpha)= \Tr_\Ff \hN \Gamma_Q =
\rhov_Q\ell^3$.  Thus we obtain the upper bound
\begin{eqnarray}\nonumber
\Tr_\Ff\, H \Gamma \!\!&\leq&\!\! \Tr_\Ff \, H^{(0)} \Gamma_Q 
+  2\pi a \rhov_Q^2  \ell^3\Big( 1 + c \Big[ \ell^{-1} 
\rhov_Q^{-1/3}+ K^2\rho_0^{4/3} \rho_Q^{-1}  s^{3/2} a^{-1/2} 
 \\ &&\qquad\qquad\qquad\qquad +E\big(R,s,K^{3/2}
\rhov_0\ell^3,K\rhov_0^{2/3},\ell \big)  \Big] \Big)  \label{up1}
\end{eqnarray}
for some constant $c>0$. 

The next step is to calculate the average particle number. 
By construction of $\Gamma$, 
\begin{equation}
\Tr_\Ff\, \hN \Gamma= \sum_\alpha \lambda_\alpha \langle 
\phi_\alpha|\hN|\phi_\alpha\rangle =  \sum_\alpha \lambda_\alpha 
\langle \psi_\alpha|\hN|\psi_\alpha\rangle= \Tr_\Ff\, \hN
\Gamma_Q\,. \label{up2}
\end{equation}

It remains to derive a lower bound on the entropy of $\Gamma$. For
this purpose, we need the following Lemma.

\begin{lem}\label{Lem1}
  Let $\Gamma$ be a density matrix on some Hilbert space, with
  eigenvalues $\lambda_\alpha\geq 0$. For $\{P_\alpha\}$ (not
  necessarily orthogonal) one-dimensional projections, let $\widehat
  \Gamma= \sum_\alpha \lambda_\alpha P_\alpha$. Then
\begin{equation}
S[\widehat \Gamma] \geq S[\Gamma] - \ln \left\|
\mbox{$\sum_\alpha$} P_\alpha\right\|\,.
\end{equation}
\end{lem}

\begin{proof}
  Using twice concavity of the logarithm,
\begin{eqnarray}\nonumber
S[\widehat\Gamma]- S[\Gamma] &=& - \sum_\alpha \lambda_\alpha \Tr
\, P_\alpha \ln\left( \lambda_\alpha^{-1} \widehat \Gamma\right)
\\ \nonumber &\geq& -\sum_\alpha \lambda_\alpha \ln \Tr\, P_\alpha
\lambda_\alpha^{-1} \widehat \Gamma \\ &\geq& - \ln \Tr \left(
\mbox{$\sum_\alpha$} P_\alpha \widehat \Gamma\right)\geq -
\ln\left\| \mbox{$\sum_\alpha$} P_\alpha\right\|\,.
\end{eqnarray}
\end{proof}

Let $\chi=\max_\alpha \|F^{N_\alpha,M_\alpha}\psi_\alpha\|^{-2}$.
Then,
\begin{equation}
\sum_\alpha |\phi_\alpha\rangle\langle\phi_\alpha| \leq \chi
\sum_\alpha F^{N_\alpha,M_\alpha}
|\psi_\alpha\rangle\langle\psi_\alpha|F^{N_\alpha,M_\alpha}\,.
\end{equation}
Note that $F^{N_\alpha,M_\alpha}$ depends on $\alpha$ only through the
particle numbers $N_\alpha$ and $M_\alpha$. Denoting the sum over a
sector of fixed $N_\alpha$ and $M_\alpha$ by $\sum'$, and using the
fact that the $\psi_\alpha$ are orthonormal, we see that
\begin{equation}
\mbox{$\sum'_\alpha$} F^{N_\alpha,M_\alpha}
|\psi_\alpha\rangle\langle\psi_\alpha|F^{N_\alpha,M_\alpha} 
\leq |F^{N_\alpha,M_\alpha}|^2\leq
1\,.
\end{equation}
This hold in every sector, and hence
\begin{equation}
\sum_\alpha |\phi_\alpha\rangle\langle\phi_\alpha| \leq \chi\,.
\end{equation}
Lemma~\ref{Lem1} thus implies that 
\begin{equation}\label{up3}
S[\Gamma]\geq S[\Gamma_Q] - \ln \chi\,, 
\end{equation}
and it remains to derive an upper bound $\chi$. Equivalently,
we need a lower bound on the norm $\|F^{N_\alpha,M_\alpha}
\psi_\alpha\|$. This can be obtained as follows.

\begin{lem}\label{chilem}
Under the same assumptions as in Lemma~\ref{enlem}, 
\begin{equation}\label{chies}
\langle\psi|\psi\rangle \geq  \left[ 1 - c\left(\frac{aR^2}{s^3}+ 
s^2 k \right)\right]_+^{\min\{N,M\}}
\left[1-c(N+M)^{8/3} (s/\ell)^5 \right]_+
\end{equation}
for some constant $c>0$. Here, $[t]_+= \max\{t , \, 0\}$ denotes the positive part.
\end{lem}

\begin{proof}
  We write $F^{N,M}(X^\upa,X^\doa)=G_N(X^\upa) G_M(X^\doa)
  H(X^\upa,X^\doa)$, where $G_N$, $G_M$ and $H$ denote the three
  different factors in (\ref{defF}). From \cite[Lemmas~1 and~3]{LSS}
  we can infer that
\begin{eqnarray}\nonumber
\langle\psi|\psi\rangle \!\!\!&=&\!\!\! \int dX^\upa\, dX^\doa \, 
D_1(X^\upa)^2 D_2(X^\doa)^2 G_N(X^\upa)^2 G_M(X^\doa)^2 H(X^\upa,X^\doa)^2  
\\ \nonumber &\geq &\!\!\! \int dX^\upa\, dX^\doa \, 
D_1(X^\upa)^2 D_2(X^\doa)^2 G_N(X^\upa)^2 H(X^\upa,X^\doa)^2   
\\ \nonumber && \times \left[ 1- c M^{8/3} \|\AaFF_{X^\upa}^{-1}\|^2 (s/\ell)^5\right]_+ 
\\ \nonumber &= &\!\!\!  \int dX^\upa\, D_1(X^\upa)^2 G_N(X^\upa)^2 
\big(\det \AaFF_{X^\upa}\big) \left[ 1- c M^{8/3} \|\AaFF_{X^\upa}^{-1}\|^2 (s/\ell)^5\right]_+ 
\,. \\  \label{normint}
\end{eqnarray}
Here, $\AaFF_{X^\upa}$ denotes the $M\times M$ matrix
\begin{equation}
\big(\AaFF_{X^\upa}\big)_{nm}= \int_{\R^3} dy\, \varphi_n^*(y) 
\varphi_m(y) \prod_{j=1}^N f(y - x^{\upa}_j)^2 \,, 
\end{equation}
with $\varphi_n$ denoting the $M$ eigenfunctions of the Dirichlet
Laplacian that constitute the Slater determinant $D_2$, and $\|\, \cdot\, \|$ stands for the
operator norm. Note that because of the factor $G_N$ the integrand in
(\ref{normint}) is only non-zero if $|x^\upa_i-x^\upa_j|\geq s$ for
all $i\neq j$. In this case, Lemma~2 in \cite{LSS} implies that
\begin{equation}
\|\id - \AaFF_{X^\upa}\| \leq c \left( \frac {aR^2}{s^3} + s^2 k \right)\,.
\end{equation}
(Again, as already mentioned after Lemma~\ref{enlem}, the factor
$n^{2/3}/\ell^2$ in the statement of \cite[Lemma~2]{LSS} has to be
replaced by $k$ in the general case considered here.) In particular,
since $0\leq \AaFF\leq \id$, this estimate implies that
\begin{equation}
\|\AaFF_{X^\upa}^{-1} \| \leq \left[ 1 - c \left( \frac {aR^2}{s^3} + s^2 k \right) \right]_+^{-1}
\end{equation}
and that
\begin{equation}
\det \AaFF_{X^\upa} \geq  \left[ 1 - c \left( \frac {aR^2}{s^3} + s^2 k \right) \right]_+^M\,.
\end{equation}
By inserting these two bounds into (\ref{normint}) and using again
\cite[Lemma~3]{LSS} to get rid of the $G_N$ in the integrand this
implies (\ref{chies}) in the case $M\leq N$. The case $N>M$ follows in
the same way, interchanging the estimates for the $X^\upa$ and
$X^\doa$-particles.
\end{proof}

To apply this lemma, we use again the fact that in the case of
interest the number of particles is at most $N+M\leq 2\, \Tr_{\Hh_1} Q
\leq c \ell^3 K^{3/2}\rhov_0$ for some constant $c>0$, and the kinetic
energy $k$ is bounded by $K \rhov_0^{2/3}$. Hence Lemma~\ref{chilem}
implies that
\begin{equation}\label{boundchi}
\frac 1\chi \geq  \left[ 1 - c\left(\frac{aR^2}{s^3}+ s^2 K
\rhov_0^{2/3}\right)\right]_+^{c\ell^3 K^{3/2}\rhov_0}
\left[1-c\ell^3 K^{4}\rhov_0 \left(s^3 \rhov_0\right)^{5/3}\right]_+
\end{equation}
for some constant $c>0$. In combination, (\ref{up1}), (\ref{up2}) and
(\ref{up3}) imply the lower bound
\begin{eqnarray}\nonumber
\Pp^\ell[\Gamma] \!\!\!&\geq& \!\!\!
\Pp_0^\ell[\Gamma_Q] - 2\pi a
\rhov_Q^2 -  \frac 1{\ell^3\beta} \ln \chi\\ \nonumber && \!\!\!\!\!\!\!\!\!\!\!\!\!\!\!\!\! - c a
\rhov_Q^2 \Big[ \ell^{-1} 
\rhov_Q^{-1/3}+ K^2 \rho_0^{4/3} \rho_Q^{-1}  s^{3/2} a^{-1/2} 
+E\big(R,s,K^{3/2}
\rhov_0\ell^3,K\rhov_0^{2/3},\ell \big) \Big] \,, \\
\label{combb}
\end{eqnarray}
with $\chi$ bounded by (\ref{boundchi}).

\subsection{Approximating Traces by Integrals}\label{sstrace}

The pressure of $\Gamma_Q$ is easy to compute:
\begin{equation}
\Pp_0^\ell[\Gamma_Q]= \frac 2{\beta\ell^3}\Tr_{\Hh_1}\, \ln
\big(1+Q \exp\big(-\beta(-\Delta-\mu)\big)\big)\,.
\end{equation}
We have to compare this quantity with the true pressure of the
non-inter\-acting gas in the thermodynamic limit,
\begin{equation}
P_0(\beta,\mu)= \frac 2{\beta} (2\pi)^{-3} \int_{\R^3} dp\, \ln
\left(1+ \exp\left(-\beta(p^2-\mu)\right)\right)\,.
\end{equation}
To this end, we note the following:

\begin{lem}\label{lemtrin}
Let $f:\R_+\mapsto\R_+$ be a monotone decreasing function, and let
$\Delta$ be the Dirichlet Laplacian on a cube of side length
$\ell$.  Then
\begin{equation}\label{trlem}
 (2\pi)^{-3} \int_{\R^3}dp\, f(p^2)\geq  \ell^{-3} 
\Tr_{\Hh_1}\, f(-\Delta) \geq
(2\pi)^{-3}\int_{\R^3}dp\, f(p^2)\left[ 1 - \frac{3\pi}{\ell |p|}
\right]\,.
\end{equation}
\end{lem}

\begin{proof}
  Note that the spectrum of $-\Delta$ is given by $[(\pi/\ell)\N]^3$.
  Considering the trace as a lower Riemann sum to the integral, we
  immediately obtain the first inequality. To obtain the second, we
  consider the trace as an upper Riemann sum to the integral over the
  region where $p_i\geq \pi/\ell$, $1\leq i\leq 3$, with $p_i$
  denoting the components of $p$. Hence
\begin{equation}\label{tr1}
(2\pi)^3 \ell^{-3} \Tr_{\Hh_1}\, f(-\Delta) \geq  \int_{\R^3}dp\,
f(p^2) - \sum_{i=1}^3 8 \int_{0\leq p_i\leq \pi/\ell}dp\, f(p^2)\,.
\end{equation}
Since $f$ is monotone decreasing, we can estimate
\begin{equation}\label{tr2}
\int_{0\leq p_i\leq \pi/\ell}dp\, f(p^2) \leq
\frac14\frac{\pi}\ell \int_{\R^2} dp\, f(p^2) = \frac 18
\frac\pi\ell \int_{\R^3} dp\, \frac 1{|p|} f(p^2)\,,
\end{equation}
where the integral in the second term is over the plane $\R^2$.
Combining (\ref{tr1}) and (\ref{tr2}) we arrive at the second
inequality in (\ref{trlem}).
\end{proof}

{}From this lemma, we immediately see that $\rhov_Q\leq \rhov_0$.
Moreover,
\begin{eqnarray}\nonumber
\Pp^\ell_0[\Gamma_Q]- P_0(\beta,\mu)\!\!\!\! &\geq& \!\!\!\!  -
\frac 2\beta (2\pi)^{-3} \int_{p^2\geq K\rhov_0^{2/3}}dp\,
\ln\left(1+z\exp(-\beta p^2)\right) \\ \nonumber && - \frac 2\beta
(2\pi)^{-3} \frac{3\pi}\ell \int_{p^2\leq K\rhov_0^{2/3}}dp\,
\frac{1}{|p|} \ln\left(1+z\exp(-\beta p^2)\right)\,.\\ \label{ppes}
\end{eqnarray}
In the integrand in the first integral, we estimate $\ln(1+x)\leq
x$ as well as $|p|\leq (2\beta)^{-1/2} \exp(\half \beta p^2)$ and
obtain
\begin{equation}
\frac 2\beta (2\pi)^{-3} \int_{p^2\geq K\rhov_0^{2/3}}dp\,
\ln\left(1+z\exp(-\beta p^2)\right)\leq \frac 1{\sqrt{2} \pi^2}
 \frac{ z}{\beta^{5/2}}
\exp\left(-\half \beta K \rhov_0^{2/3}\right)\,.
\end{equation}
The second term on the right side of (\ref{ppes}) is bounded from below by
\begin{equation}
- \frac{3}{4\pi^2} \frac 1{\beta^2\ell}\int_{\R^3}dp\,
\frac{1}{|p|} \ln\left(1+z\exp(-p^2)\right)\equiv - \frac{3}{4\pi^2}
\frac 1{\beta^2\ell} g(z)\,,
\end{equation}
and therefore
\begin{equation}\label{combi}
\Pp^\ell_0[\Gamma_Q]\geq  P_0(\beta,\mu) - \frac 1{\pi^2
\beta^{5/2}} \left[ \frac{3\beta^{1/2}}{4 \ell} g(z) +
\frac{z}{\sqrt 2 } \exp\left(-\half \beta K
\rhov_0^{2/3}\right)\right]\,.
\end{equation}
By combining (\ref{combi}) with (\ref{combb}) and the estimate $\rhov_Q\leq
\rhov_0$, we thus obtain, for some constant $c>0$,
\begin{eqnarray}\nonumber
P^\ell(\beta,\mu)\!\!\!&\geq&\!\!\!  P_0(\beta,\mu) - 2\pi a
\rhov_0^2- \frac 1{\ell^3\beta} \ln \chi \\ 
\nonumber && \!\!\!\!\!\!\!\!\! - c a
\rhov_0^2 \Big[ \ell^{-1} 
\rhov_0^{-1/3}+ K^2\rho_0^{1/3} s^{3/2} a^{-1/2} 
+E\big(R,s,K^{3/2}
\rhov_0\ell^3,K\rhov_0^{2/3},\ell \big) \Big]
\\  &&
\!\!\!\!\!\!\!\!\! - \frac 1{\pi^2 \beta^{5/2}} \left[ \frac{3\beta^{1/2}}{4
\ell} g(z) + \frac{z}{\sqrt 2 } \exp\left(-\half \beta K
\rhov_0^{2/3}\right)\right]\,, \label{secl}
\end{eqnarray}
with $\chi$ bounded by (\ref{boundchi}).

\subsection{Final Result}\label{ssfinal}

We are still free to choose $\ell$, $R$, $s$ and $K$. We choose
\begin{equation}
R=a(a^3\rhov_0)^{-1/81}\ , \ s=a(a^3\rhov_0)^{-10/81}\ , \
\ell=\rhov_0^{-1/3}(a^3\rhov_0)^{-28/81}
\end{equation}
and $K=(a^3\rhov_0)^{-\eps/12}$ for some $\eps>0$. 
With this choice the first term in square brackets in (\ref{secl}) is bounded by
\begin{equation}
\ell^{-1} 
\rhov_0^{-1/3}+ K^2\rho_0^{1/3} s^{3/2} a^{-1/2} 
+E\big(R,s,K^{3/2}
\rhov_0\ell^3,K\rhov_0^{2/3},\ell \big) \leq c
 \big (a\rhov_0^{1/3}\big)^{1/27-\eps}
\end{equation}
for some $c>0$ and $a^3\rho_0$ small. 
Moreover,
\begin{equation}
\frac 1{\ell^3\beta} \ln \chi \leq \const \frac 1{\beta\rhov_0^{2/3}} 
a\rhov_0^2 \big (a\rhov_0^{1/3}\big)^{1/27-3\eps/8}\,.
\end{equation}
Note that $\beta\rhov_0^{2/3}$ is a monotone increasing function of
$z$ and, in particular, $1/(\beta\rhov_0^{2/3})$ is bounded for
bounded $1/z$.

The first term in the last line of (\ref{secl}) is
\begin{equation}\label{secl2}
\frac {1}{\beta^2\ell} g(z) = a\rhov_0^2 \frac{g(z)}
{\big(\beta\rhov_0^{2/3}\big)^2} \big (a\rhov_0^{1/3}\big)^{1/27}\,.
\end{equation}
Now $\beta\rhov_0^{2/3} \sim \ln(z)$ for large $z$, and also
$g(z)\sim \ln(z)$ for large $z$. Hence the fraction in (\ref{secl2}) is
uniformly bounded in $z$ for bounded $1/z$. The remaining term in
(\ref{secl}) is
\begin{equation}\label{secl3}
\rhov_0^{5/3} \frac 1{\big(\beta\rhov_0^{2/3}\big)^{5/2}} 
\left[ \exp\left(-\half \beta\rhov_0^{2/3}+ K^{-1}\ln z\right)\right]^K\,.
\end{equation}
For large $K$, the term in square brackets is strictly less than one,
again uniformly in $z$. Hence we see that the expression (\ref{secl3})
is exponentially small for small $a^3\rhov_0$ and, in particular,
bounded by $\const \rhov_0^{5/3} (a^3\rhov_0)^p$ for any exponent $p$,
with a constant that depends on $p$ (and $\eps$), of course, but not
on $z$ (for bounded $1/z$).

To summarize, we have thus shown that with the choice of parameters as
above, (\ref{secl}) gives, for any $\eps>0$,
\begin{equation}
P^\ell(\beta,\mu)\geq P_0(\beta,\mu) - 2\pi a \rhov_0^2 \left( 1 +
C_\eps(z) \big (a\rhov_0^{1/3}\big)^{1/27-\eps}\right)\,,
\end{equation}
with some constant $C_\eps(z)$ that is uniformly bounded in $z$ for
bounded $1/z$. To complete the estimate, we have to insert this bound
into (\ref{boxbound}). Thus we still have to estimate
\begin{equation}\label{347}
\frac{R_0}
\ell P_0(\beta,\mu)= \frac{R_0}{a} a\rhov_0^2 
\big (a\rhov_0^{1/3}\big)^{28/81} \big(\beta\rhov_0^{2/3}\big)^{-5/2} 
P_0(1,\beta\mu)\,.
\end{equation}
Now $P_0(1,\beta\mu)\sim (\ln z)^{5/2}$ for large $z$, and therefore
$(\beta\rhov_0^{2/3})^{-5/2} P_0(1,\beta\mu)$ is uniformly bounded for
bounded $1/z$. Altogether, this implies that, for some constant
$C_\eps(z)$ uniformly bounded for bounded $1/z$,
\begin{equation}
P(\beta,\mu)\geq P_0(\beta,\mu) - 2\pi a \rhov_0^2 \left( 1 +
C_\eps(z) \big (a\rhov_0^{1/3}\big)^{1/27-\eps}\right)\,.
\end{equation}
This finishes the proof of the lower bound.

\section{Upper Bound to the Pressure}\label{UB}

In maximizing the pressure functional we can restrict ourselves to
density matrices that do not mix particle numbers. More precisely, if
$Q_N^\updownarrow$ denotes the projection onto the sector of $N$
particles in $\Ff_\updownarrow$, then $\Pp^L[\Gamma]\leq
\Pp^L[\widehat \Gamma]$, with $\widehat\Gamma=\sum_{N,M} Q^\upa_N
Q^\doa_M \Gamma Q^\upa_N Q^\doa_M$. This follows from the fact that
the entropy is non-decreasing under this transformation
\cite[2.1,11.4]{Thirring}.  Hence we can assume that
$\Gamma=\bigoplus_{N,M}\Gamma_{N,M}$, with $\Gamma_{N,M}$ (not
normalized) density matrices for $N$ $\upa$-particles and $M$
$\doa$-particles, respectively.  For simplicity, we may also assume
that $\Gamma$ is symmetric with respect to exchange of $\upa$ and
$\doa$. This is certainly no restriction in the case considered here,
and leads to simpler expressions by shortening some of the formulas.

\subsection{Lower Bound to the Hamiltonian}\label{sslow}

We start with a lemma which is essentially a generalization of a result
by Dyson \cite{dyson} to bound the hard potential $v$ from below by a
soft potential $U$, at the expense of some kinetic energy. In the
following, $\widehat f$ denotes the Fourier transform of a function
$f$. We use the convention $\widehat f(p)=(2\pi)^{-3/2} \int dx\, f(x)
e^{-ipx}$.

\begin{lem}\label{dyson}
  Let $\chi$ be a radial function, $0\leq \chi\leq 1$, with
  $h\equiv\widehat{1-\chi} \in L^1(\R^3)\cap L^\infty(\R^3)$. For $R>R_0$,
  ´let
\begin{equation}\label{deffr}
f_R(x)= \sup_{|y|\leq R} | h(x-y) - h(x) |\,,
\end{equation}
and
\begin{equation}\label{defwr}
w_R(x)= \frac{2}{\pi^2}  f_R(x) \int_{\R^3} dy \, f_R(y)\, .
\end{equation}
Let $U$ be a positive, radial function, supported in the annulus
$R_0\leq |x|\leq R$, with $\int_{\R^3} dx\, U(x) = 4\pi$. If
$y_1,\dots,y_M$ denotes a set of $M$ points in $\R^3$, with
$|y_k-y_l|\geq 2R$ for all $k\neq l$, then, for any $\eps>0$,
\begin{equation}
 -\nabla \chi(p)^2 \nabla + \half \sum_{k=1}^M v(x-y_k) \geq
\sum_{k=1}^M \left( (1-\eps) a U(x-y_k) - \frac a\eps w_R(x-y_k)
\right)
\end{equation}
in the sense of quadratic forms.  Here, $\chi(p)$ stands for the
multiplication operator in momentum space. This operator inequality
holds for all functions in $H^1(\R^3)$ and therefore, in particular,
for functions supported in the cube $\Lambda_L$.
\end{lem}

The proof of this lemma can be found in \cite[Lemma~4 and
Cor.~1]{LSS}.  Note that, by construction, either $w_R\in
L^1(\R^3)\cap L^\infty(\R^3)$, or $w_R\equiv \infty$ identically.

Lemma~\ref{dyson} implies the following lower bound to the Hamiltonians
(\ref{ham2}):
\begin{eqnarray}\nonumber
H_{N,M} &\geq & \sum_{i=1}^N \left[-
\nabla_{i}^\upa\left(1-\chi(p_{i}^\upa)^2\right)
\nabla_{i}^\upa + 
W_{X^\doa}(x^\upa_i) \right] \\  && + \sum_{k=1}^M
\left[ -\nabla_{k}^\doa\left(1-\chi(p_{k}^\doa)^2\right)
\nabla_{k}^\doa +
 W_{X^\upa}(x^\doa_k) \right]\,.  \label{hamlow}
\end{eqnarray}
Here, $W_Y$ is the potential
\begin{equation}\label{defwy}
W_Y(x) = \sum_{\{ k\, : \, y_k \in \widetilde Y_R\}} \left(
(1-\eps) a U(x-y_j) - \frac a\eps w_R(x-y_j) \right)\,,
\end{equation}
where $Y=(y_1,\dots,y_M)$ is any set of coordinates in $\Lambda_L$ and
$\widetilde Y_R \subset Y$ is the subset of $y_j$'s whose distance to
the nearest neighbor in $Y$ is at least $2R$. We neglect the
interaction with $y_j$'s that are not in the set $\widetilde Y_R$,
which can only lower the energy. Note that also the interaction terms
among particles of equal spin are dropped for a lower bound.

The expectation value of the potentials $W_{X^\upa}$ and $W_{X^\doa}$
can be written in the following convenient way.  For
$\Gamma=\bigoplus_{N,M} \Gamma_{N,M}$ a density matrix on
$\Ff_\upa\otimes \Ff_\doa$, and $X^\doa =(x^\doa_1,\dots,x^\doa_M)$
some fixed coordinates of the spin-down particles, let the operator
$\Gamma_N^{X^\doa}$ be defined through the integral kernel
$\Gamma_N^{X^\doa}(X^\upa,Y^\upa)=
\Gamma_{N,M}(X^\upa,X^\doa,Y^\upa,X^\doa)$. Since $\Gamma_{N,M}$ is a
trace class operator, this expression is well defined for almost every
$X^\doa$ by an eigenfunction expansion of $\Gamma_{N,M}$. The same is
true for $n(X^\doa)$, given by $n(X^\doa)=
\Tr_{\Ff_\upa}[\bigoplus_N\Gamma_N^{X^\doa}]$. Note that $n(X^\doa)$
is the probability density for having exactly $M$ spin-down particles
at the positions $X^\doa$. In case $n(X^\doa)>0$, let
\begin{equation}\label{gexp}
\Gamma_\upa^{X^\doa}= n(X^\doa)^{-1} \bigoplus_N \Gamma_N^{X^\doa}\,.
\end{equation}
This defines $\Gamma_\upa^{X^\doa}$ only if $n(X^\doa)$
is non-zero; only in this case it will be used below, however. Note
that $\Gamma_\upa^{X^\doa}$ is a density matrix on $\Ff_\upa$, which
can be interpreted as the density matrix of the $\upa$-particles for a
fixed configuration of the $\doa$-particles. If $\gamma_\upa^{X^\doa}$
denotes the reduced one-particle density matrix of
$\Gamma_\upa^{X^\doa}$, then the expectation value of the potential
$W_{X^\doa}$ in the state $\Gamma$ can be written as
\begin{equation}\label{bound47}
\intsum dX^\doa\, n(X^\doa)\, \Tr_{\Hh_1}
\left[ W_{X^\doa} \gamma_\upa^{X^\doa}\right] \,,
\end{equation}
where we introduced the short hand notation $\intsum dX^\doa \equiv
\sum_M \int dx^\doa_1\cdots dx^\doa_M$. Note that $\intsum dX^\doa\,
n(X^\doa) = 1$, and $\intsum dX^\doa\, n(X^\doa)\, \Gamma_\upa^{X^\doa}=
\Gamma_\upa \equiv \Tr_{\Ff_\doa}\, \Gamma$.

Under the assumption that the density matrix is symmetric with respect
to exchange of $\upa$ and $\doa$-particles, the lower bound to the
Hamiltonian in (\ref{hamlow}) can thus be written as follows:
\begin{equation}\label{implie}
\Tr_{\Ff} \, H \Gamma \geq 2\, \intsum dX^\doa \, n(X^\doa)\, 
\Tr_{\Hh_1} \left[ \big(-\nabla(1-\chi(p)^2)\nabla + 
W_{X^\doa} \big)\gamma_\upa^{X^\doa}\right]\,.
\end{equation}
Note that the kinetic energy term does not depend on $X^\doa$ and,
therefore, the integration affects only $\gamma_\upa^{X^\doa}$. Note
also that $\intsum dX^\doa\, n(X^\doa)\, \gamma_\upa^{X^\doa}=
\gamma_\upa$, the reduced one-particle density matrix for the
$\upa$-particles.

For reasons which will become clear later, we find it convenient not
to use up all the kinetic energy in the bound (\ref{implie}), however.
More precisely, we pick some $0<\delta<1$ and $0<\kappa<1$ and
write $-\Delta$ as
\begin{equation}\label{thr}
-\Delta = -\delta \Delta - (1-\delta) (1-\kappa) \nabla \chi(p)^2 \nabla 
+ h^\chi\,,
\end{equation}
with
\begin{equation}\label{defhchi}
h^\chi = - (1-\delta)\nabla \big(1-(1-\kappa)\chi(p)^2\big) \nabla\,.
\end{equation}
Applying the above estimate only to the second term on the right side
of (\ref{thr}) and using positivity of the interaction potential $v$,
we obtain that
\begin{eqnarray}\nonumber
\!\!\!\!\!\Tr_{\Ff} \, H \Gamma \!\!&\geq&\!\! 2\, \Tr_{\Hh_1} \left[ \big( -\delta \Delta 
+ h^\chi\big) \gamma_\upa\right] \\ && \!\!+ 2 (1-\delta)(1-\kappa) 
\, \intsum dX^\doa \, n(X^\doa)\, 
\Tr_{\Hh_1} \left[ 
W_{X^\doa} \gamma_\upa^{X^\doa}\right]\,. \label{finsub}
\end{eqnarray}
Eq. (\ref{finsub}) is the final result of this subsection. To estimate
this expression, we will show that, for any fixed $X^\doa$, the
one-particle density matrix $\gamma_\upa^{X^\doa}$ is close to the
corresponding expression for non-interacting particles (which does not
depend on $X^\doa$, of course). We do this in Subsection~\ref{apse}
below.

It remains to choose $\chi$.  Let $\eta:\R^3\to \R_+$ be a smooth radial
function with $\eta(x)=0$ for $|x|\leq 1$, $\eta(x)=1$ for $|x|\geq 2$, and
$0\leq \eta(x)\leq 1$ in-between. For some $s>0$ we choose
\begin{equation}\label{defchi}
\chi(p)=\eta(s p)\,.
\end{equation}
The potential $W_Y$ then depends on $\eps$, $a$, $R$ and $s$. Note
that with this choice of $\chi$ the corresponding $h=\widehat{1-\chi}$
is a smooth function of rapid decay and hence, by simple scaling, the
corresponding potential $w_R$ defined in (\ref{defwr}) satisfies, for
$R\leq \const s$,
\begin{equation}\label{intw}
\|w_R\|_\infty \leq \const \frac{R^2}{s^5} \ \quad {\rm and\quad }
\|w_R\|_1 \leq \const \frac {R^2}{s^2}
\end{equation}
for some constants depending only on $\eta$, which is fixed once and for
all. Moreover, if $|y_k-y_l|\geq 2R$ for all $k\neq l$, then
\begin{equation}\label{normw}
\sum_{k=1}^M  w_R(x-y_k) \leq \const \frac 1{R s^2}
\end{equation}
independently of $x$ and $M$.

We are also still free to choose the potential $U$ in Lemma~\ref{dyson}. We choose 
\begin{equation}
U(x)=\left\{ \begin{array}{ll} 3\left(R^3-R_0^3\right)^{-1} 
& {\rm for\ } R_0\leq |x|\leq R, \\ 0 & {\rm otherwise}\,. 
\end{array} \right.
\end{equation}
We then have the estimate
\begin{equation}\label{normW}
\| W_Y \|_\infty \leq \max\left\{ \frac {3a}{R^3-R_0^3}\, , \, 
\const \frac a{\eps R s^2}\right\}\,,
\end{equation}
independently of $Y$. 

\subsection{Improved Subadditivity of Entropy}\label{ssl2}

For $\Gamma$ a density matrix on $\Ff_\upa\otimes \Ff_\doa$, let
$\Gamma_\upa = \Tr_{\Ff_\doa}\, \Gamma$ and $\Gamma_\doa =
\Tr_{\Ff_\upa}\, \Gamma$ be the density matrices of the subsystems of
$\upa$ and $\doa$-particles, respectively. It is well known that the
entropy $S[\Gamma]$ is subadditive (see, e.g.,
\cite[Ineq.~(2.2,13)]{Thirring}, i.e.,
\begin{equation}\label{ineq1}
S[\Gamma]\leq S[\Gamma_\upa]+ S[\Gamma_\doa]\,,
\end{equation}
where the entropies on the right side are defined by taking the trace
only over $\Ff_\upa$ and $\Ff_\doa$, respectively. Moreover, if
$\gamo_\updownarrow$ denotes the reduced one-particle density matrix
of $\Gamma_\updownarrow$, then
\begin{equation}\label{ineq2}
S[\Gamma_\updownarrow]\leq \SsFF[\gamo_\updownarrow]\,,
\end{equation}
where $\SsFF[\gamo]=\Tr_{\Hh_1}\left(-\gamo \ln
  \gamo-(1-\gamo)\ln(1-\gamo)\right)$ is given as in (\ref{defss}) 
\cite[Ineq.~(2.5,18.5)]{Thirring}.  Note that both (\ref{ineq1}) and
(\ref{ineq2}) are equalities if $\Gamma$ is the grand-canonical Gibbs
density matrix of a non-interacting system.

We are going to need the following refinement of subadditivity of
entropy. Its proof is given in \cite[Cor.~4]{LSent}.

\begin{lem}\label{lemsub}
  Let $\Gamma=\bigoplus_{N,M} \Gamma_{N,M}$ be a density matrix on
  $\Ff_\upa\otimes \Ff_\doa$. With the notation introduced in
  Subsect.~\ref{sslow},
\begin{equation}\label{gensub}
S[\Gamma]\leq S[\Gamma_\doa] + \intsum dX^\doa \, n(X^\doa)\, 
S[\Gamma_\upa^{X^\doa}]\,.
\end{equation}
\end{lem}

Note that $\intsum dX^\doa\, n(X^\doa) = 1$, and $\intsum dX^\doa\,
n(X^\doa) \Gamma_\upa^{X^\doa}= \Gamma_\upa$. Hence, by con\-cavity of
$S[\Gamma_\upa]$, inequality (\ref{gensub}) is stronger than the usual
subadditivity of entropy in (\ref{ineq1}). The last term on the right
side of (\ref{gensub}) is the average entropy of the $\upa$-particles
for fixed $\doa$-particles, whereas $S[\Gamma_\upa]$ is the entropy of
the state of $\upa$-particles averaged over all configurations of the
$\doa$-particles.

\subsection{A Priori Bounds on the One-Particle Density Matrix}\label{apse}

In this subsection we will show that the one-particle density matrices
for fixed $\doa$-particles, $\gamma_\upa^{X^\doa}$, are close to the
corresponding expression for non-interacting particles, provided the
state $\Gamma$ that defines them has a variational pressure
$\Pp^L[\Gamma]$ close to the true pressure $P^L(\beta,\mu)$ of the
system, i.e., it is an approximate maximizer of the pressure
functional (\ref{func}) in a sense to be made precise below. We call
such a bound an {\it a priori} bound. 

This subsection is split into four parts. In Part~1, we will
derive a bound of the sort needed.  This bound will not be uniform in the
fugacity $z$, however, and will be useless for very large $z$ (when
the system is close to its ground state). For large $z$, however, we
will use a different method to obtain a similar bound, by comparing
the state with the ground state of the non-interacting system. We do
this in Part~3. Before that, we use the method of Part~1 to
compare the non-interacting system with different boundary conditions
(Part~2). Some calculations are easier to do with periodic
boundary conditions than with Dirichlet, hence the usefulness of this
estimate. Finally, we give a summary of the result of this subsection
in Part~4.

\subsubsection{General a priori bound}\label{sss1}

Using the fact that $v\geq 0$, we can infer from Lemma~\ref{lemsub}
and (\ref{ineq2}) that
\begin{eqnarray}\nonumber
\!\!\!\!\!\!\!\!\!\!\!
-L^3 \Pp^L[\Gamma]\!\!\! &\geq&\!\!\! \Tr_{\Hh_1} \left[\left(
-\Delta- \mu\right)\gamo_\doa\right] -  \frac 1\beta
\SsFF[\gamo_\doa] \\ &&\!\!\! + \intsum dX^\doa\, n(X^\doa) \left(
\Tr_{\Hh_1} \left[\left( -\Delta-
\mu\right)\gamo_\upa^{X^\doa}\right] - \frac 1\beta
\SsFF\big[\gamo_\upa^{X^\doa}\big]\right)\, .\label{apr1}
\end{eqnarray}
The term in the first line on the right side of (\ref{apr1}) is
bounded from below by $-\half L^3 P_0^L(\beta,\mu)$ because of
(\ref{onepf}). The same is true for the term in the last line, but we
will need a refinement of this inequality, given in Lemma~\ref{oplem}
below.

For $\Gamma$ an approximate maximizer of $\Pp^L[\Gamma]$, we have an
upper bound on the left side of (\ref{apr1}), derived in the previous
section, and therefore this yields an upper bound on the last
expression on the right side of (\ref{apr1}). This bound can be used
to get information on the one-particle density matrices
$\gamma_\upa^{X^\doa}$. We need the following lemma.

\begin{lem}\label{oplem}
  Let $h$ be a self-adjoint operator on a Hilbert space $\Hh$, such
  that $e^{-h}$ is trace class. For $\gamo$ a fermionic one-particle
  density matrix (i.e., a trace class operator on $\Hh$ with $0\leq
  \gamo\leq 1$), define the functional
\begin{equation}\label{eh}
\Ecal_h[\gamo]=\Tr\, h\gamo - \SsFF[\gamo]\,.
\end{equation}
Let $\gamo_h=(1+e^h)^{-1}$ be its minimizer, and
$f(h)=\Ecal_h[\gamo_h]=-\Tr\, \ln(1+e^{-h})$. Then, for any $\gamo$,
\begin{equation}\label{lemapr}
\Ecal_h[\gamo]\geq f(h) + 2\, \Tr (\gamo-\gamo_h)^2\, ,
\end{equation}
and also
\begin{equation}\label{lemapr2}
\Ecal_h[\gamo]\geq f(h)  +\frac 12 \frac{\left| \Tr (\gamo-\gamo_h)
\right|^2}{\left|\Tr (\gamo-\gamo_h)
\right|+ \Tr\, \gamo_h }\,.
\end{equation}
\end{lem}

Note that this lemma implies, in particular, that $\gamma\to \gamma_h$
in trace class norm, if $\Ecal_h[\gamma]\to f(h)$. Eq.~(\ref{lemapr})
implies convergence in Hilbert-Schmidt norm, but because of
(\ref{lemapr2}) also the traces converge, and therefore the
convergence is in trace class norm (see \cite{wehrl3,S79} or
Ineq.~(\ref{concl}) below).

\begin{proof}
We write
\begin{equation}
\Ecal_h[\gamo]-f(h) =\Tr\, g(\gamo,\gamo_h)\, ,
\end{equation}
with
\begin{equation}
g(\gamo,\gamo_h)=\gamo \ln \gamo - \gamo\ln \gamo_h +
(1-\gamo) \ln (1-\gamo) -(1-\gamo) \ln (1-\gamo_h)\,.
\end{equation}
The function $g$ has the integral representation (for $0\leq x,y\leq
1$)
\begin{equation}\label{interep}
g(x,y)=\int_y^x dz\, (x-z) \left(\frac 1z +\frac 1{1-z}\right)\,.
\end{equation}
Using that $1/z+1/(1-z)\geq 4$ in the integrand, we obtain the lower
bound $g(x,y)\geq 2(x-y)^2$. Hence, by Klein's inequality
\cite[Ineq.~(2.1,7.5)]{Thirring}, $\Tr\, g(\gamo,\gamo_h) \geq 2\,
\Tr(\gamo-\gamo_h)^2$, and (\ref{lemapr}) follows. Moreover,
estimating $1/z+1/(1-z)\geq 1/z\geq 1/\max\{x,y\}\geq 1/(|x-y|+y)$ in
the integrand in (\ref{interep}), we obtain
\begin{equation}\label{bb}
g(x,y)\geq \frac 12 \frac{(x-y)^2}{|x-y|+y} =  2 \sup_{0<b<1}
\left[ b(1-b) |x-y| - b^2 y\right]\,.
\end{equation}
Hence, again by Klein's inequality,
\begin{equation}
\Tr \, g(\gamo,\gamo_h) \geq 2  b(1-b)  |\Tr (\gamo-\gamo_h) | 
- 2 b^2 \Tr\, \gamo_h 
\end{equation}
for any $0<b<1$. Taking the supremum over $b$ yields (\ref{lemapr2}).
\end{proof}

We now apply this lemma to (\ref{apr1}), with $h=\beta(-\Delta-\mu)$.
Note that $f(h)=- \half \beta L^3 P_0^L(\beta,\mu)$ in this case. Let
$\gamo_0=\gamo_h=(1+z^{-1} \exp(-\beta \Delta))^{-1}$ be the minimizer
of (\ref{eh}). We can infer from (\ref{lemapr}) and (\ref{apr1}) that
\begin{equation}
\intsum dX^\doa\, n(X^\doa)\, \Tr\big(\gamo_\upa^{X^\doa}-\gamo_0\big)^2
\leq \half\beta L^3 \left(  P_0^L(\beta,\mu) - \Pp^L[\Gamma]
\right)\,.
\end{equation}
If $L$ is large, $a^3\rhov_0$ small, and $z=e^{\beta \mu}$ is bounded
away from zero, the lower bound to the pressure derived in the previous section
shows that we can restrict our attention to density matrices $\Gamma$ with
$\Pp^L[\Gamma]\geq P_0^L(\beta,\mu)- C a\rhov_0^2$, for some constant
$C>2\pi$. Hence, for such a $\Gamma$,
\begin{equation}\label{2nor}
\intsum dX^\doa\, n(X^\doa)\, \Tr\big(\gamo_\upa^{X^\doa}-\gamo_0\big)^2
\leq \half C\beta L^3 a\rhov_0^2\,.
\end{equation}
Using (\ref{lemapr2}) instead of (\ref{lemapr}), we obtain in the same
way
\begin{equation}\label{tres}
\intsum dX^\doa\, n(X^\doa)\,
 \frac{\left| \Tr \big(\gamo_\upa^{X^\doa}-\gamo_0\big)
\right|^2}{\left| \Tr \big(\gamo_\upa^{X^\doa}-\gamo_0\big)
\right|+ \Tr\, \gamo_0}
 \leq 2 C\beta L^3
a\rhov_0^2\,.
\end{equation}
By using convexity of the map $x\mapsto x^2/(x+1)$, as well as the fact
that, by Lemma~\ref{lemtrin}, $\Tr\, \gamo_0\leq \half L^3 \rhov_0$,
(\ref{tres}) implies that
\begin{equation}\label{tres2}
\intsum dX^\doa\, n(X^\doa)\,
 \left| \Tr \big(\gamo_\upa^{X^\doa}-\gamo_0\big)
\right| \leq L^3\rhov_0 \sqrt{ C a \rhov_0 \beta} 
\left( 1+ \sqrt{4 C a \rhov_0 \beta}\right)\,. 
\end{equation}
We thus have an upper bound on both the average Hilbert-Schmidt norm of the
difference of $\gamo_\upa^{X^\doa}$ and $\gamma_0$ and the average
difference of their trace. We can thus obtain a bound on the average
trace norm of their difference, which will be needed in the next
subsection.

Let $a$ and $b$ be two positive trace class operators, let $P$ be a
projection with finite rank, and set $Q=1-P$. With $\|\, \cdot \|_p =
(\Tr [|\, \cdot\, |^p])^{1/p}$ denoting the Schatten $p$-norm, we
have
\begin{eqnarray}\nonumber
\!\!\!\!\!\!\! \|a-b\|_1 \!\!&\leq&\!\! \|(a-b)P \|_1+ \|a Q\|_1 + \|b Q\|_1 \\ 
&\leq&\!\! \|P\|_2 \|a-b\|_2 + \|a\|_1^{1/2} \|QaQ\|_1^{1/2} +  
\|b\|_1^{1/2} \|QbQ\|_1^{1/2}\,.
\end{eqnarray}  
The trace norm of $a$ can be estimated by $\|a\|_1\leq \|b\|_1 + |\Tr\,
(a-b) |$. Moreover,
\begin{eqnarray}\nonumber
\|QaQ\|_1 &=& \Tr\, aQ = \Tr\left[ bQ + (a-b) + (a-b)P\right] \\ 
&\leq & \|QbQ\|_1 + |\Tr(a-b)| + \|P\|_2 \|a-b\|_2\,.
\end{eqnarray}
In conclusion, we thus obtain that
\begin{eqnarray}\nonumber
\|a-b\|_1 \leq \|P\|_2 \|a-b\|_2 \!\!\!\! &+& \!\!\!\!2 
\big( \|b\|_1 + |\Tr\, (a-b) |\big)^{1/2} \\ \nonumber &&\!\!\!\!
\times  \big( \|QbQ\|_1 + |\Tr(a-b)| + \|P\|_2 \|a-b\|_2\big)^{1/2}\,.
\\ \label{concl}
\end{eqnarray}

We apply this inequality, with $a=\gamo_\upa^{X^\doa}$ and
$b=\gamo_0$, using the estimates (\ref{2nor}) and (\ref{tres2}). We
choose $P$ to be the projection onto the subspace of
$\Hh_1=L^2(\Lambda_L;\C)$ where $-\Delta\leq K \rhov_0^{2/3}$ for some
$K>0$. Using Lemma~\ref{lemtrin}, we have $\|b\|_1= \Tr\,\gamma_0 \leq \half L^3
\rhov_0$. Moreover, again by Lemma~\ref{lemtrin}, we can estimate 
\begin{equation}
\|P\|_2^2 = \Tr\, P \leq \frac {L^3}{(2\pi)^{3}} \int dp\, 
\theta( K\rhov_0^{2/3} - p^2) = L^3 \rhov_0 \frac{K^{3/2}}{6\pi^2}
\end{equation}
and also 
\begin{eqnarray}\nonumber
\|QbQ\|_1\!\!&=&\!\!\Tr\, Qb \leq z\, \Tr\, e^{\beta \Delta }\theta(-\Delta-K\rhov_0^{2/3}) 
\\ \nonumber &\leq&\!\! z \exp\big(-\half\beta K \rhov_0^{2/3}\big) \Tr\, e^{\half\beta\Delta} 
\leq   L^3\rhov_0  \frac{z \exp\big(-\half\beta K \rhov_0^{2/3}\big)}
{(2\pi\beta)^{3/2} \rhov_0}\,. \\ \label{cond}
\end{eqnarray}
Similarly to the discussion of the term (\ref{secl3}) in the
calculation of the lower bound, this last fraction is exponentially
small in $a^3\rhov_0$ if we choose $K=(a^3\rhov_0)^{-\nu}$ for some
$\nu>0$, uniformly in $z$ for bounded $1/z$. By using (\ref{2nor}),
(\ref{tres2}) and (\ref{concl})--(\ref{cond}), as well as the Schwarz
inequality for the integration over $X^\doa$, we infer that, for small
$a^3\rhov$ and $z$ bounded away from zero,
\begin{equation}\label{tres3}
\intsum dX^\doa\, n(X^\doa)\,
 \left\| \gamo_\upa^{X^\doa}-\gamo_0 \right\|_1 
\leq C_\nu L^3\rhov_0 \big (a\rhov_0^{1/3}\big)^{1/4-3\vu/8} 
\big(\beta\rhov_0^{2/3}\big)^{1/4}
\end{equation}
for some constant $C_\nu$ depending on $\nu$. 

\subsubsection{Comparing different boundary conditions}\label{sss2}

In the following, it will be convenient to compare
$\gamo_\upa^{X^\doa}$ not with $\gamo_0$ but rather with $\gamo_\per$,
which is the minimizer of (\ref{eh}) with $h=\beta(-\Delta_\per-\mu)$,
$\Delta_\per$ denoting the Laplacian with {\it periodic boundary
  conditions} on the cube $\Lambda_L$. Note that $\gamo_\per$ has a
strictly constant density. The density matrices $\gamo_0$ and
$\gamo_\per$ agree in the thermodynamic limit, however.  This can be
seen as follows. Since the quadratic form domain of $\Delta$ is
included in the quadratic form domain of $\Delta_\per$, we can use
$\gamo_0$ as a trial density matrix of $\Ecal_h$ with
$h=\beta(-\Delta_\per-\mu)$. Since the pressure is independent of
boundary conditions in the thermodynamic limit,
\begin{equation}
\lim_{L\to\infty} \frac 1{L^3} \left( f\big(\beta(-\Delta_\per-\mu)\big) 
- f\big(\beta(-\Delta-\mu)\big)\right)=0 \,.
\end{equation}
Thus Lemma~\ref{oplem} together with (\ref{concl}) implies that, for
fixed $\beta$ and $\mu$,
\begin{equation}\label{thermo}
\lim_{L\to\infty} \frac 1{L^3} \left\| \gamo_0 - \gamo_\per\right\|_1 = 0 \,.
\end{equation}

\subsubsection{A bound uniform in the fugacity}\label{sss3}
 
The estimate (\ref{tres3}) is not uniform in the fugacity
$z=e^{\beta\mu}$. In fact, $\beta\rhov_0^{2/3}$ grows like $\ln (z)$
for large $z$. Note that large $z$ corresponds to the low-temperature
limit where the system approaches its ground state. Hence, for large
$z$, we will compare $\gamo_\upa^{X^\doa}$ with the Fermi sea
corresponding to the ground state, namely $P_\mu\equiv
\theta(\mu+\Delta)$. More precisely, we are going to use (\ref{tres3})
only in the case when $\beta\rhov_0^{2/3} \leq (a^3\rhov_0)^{-1/9}$.
For the case of larger $z$, where $\beta\rhov_0^{2/3} >
(a^3\rhov_0)^{-1/9}$, we now derive a separate bound.

We start with the following estimate. Let $Q_\mu=1-P_\mu$ and
$e(\mu)=\Tr\, (-\Delta-\mu)P_\mu$. For non-negative numbers $r,s\geq 0$, and for any
operator $\gamma$ with $0\leq \gamma\leq 1$,
\begin{eqnarray}\nonumber
&& \Tr\, (-\Delta-\mu+r P_\mu - s Q_\mu)\gamma \\ \nonumber &&\geq  
\Tr\, (-\Delta-\mu+r P_\mu - s Q_\mu)\theta(\mu+\Delta-rP_\mu+sQ_\mu) \\ 
&&= e(\mu-r) + e(\mu+s) -e(\mu) + s\, \Tr\, P_\mu\,.
\end{eqnarray}
Hence
\begin{eqnarray}\nonumber
&& \Tr\, (-\Delta-\mu)\gamma - e(\mu) \\ \nonumber && 
\geq \big[ e(\mu-r)-e(\mu) -r\, \Tr P_\mu\big]  + 
\big[  e(\mu+s) -e(\mu) + s\, \Tr\, P_\mu\big] \\ && \quad 
+ r\, \Tr\, P_\mu (1-\gamma) + s\, \Tr\, (1-P_\mu) \gamma\, . \label{442}
\end{eqnarray}
Note that, in the thermodynamic limit (and for $\mu\geq 0$), 
\begin{eqnarray}\nonumber
\lim_{L\to\infty} \frac 1{L^3} \big[ e(\mu-r)-e(\mu) -r\, 
\Tr P_\mu\big] \!\!\!\!&=&\!\!\!\! \frac 1{6\pi^2} 
\left[ -\frac 25 [\mu-r]_+^{5/2} + \frac 25 \mu^{5/2} - 
r \mu^{3/2} \right] \\&  \geq& \!\!\!\!  - \frac 1{8\pi^2} 
\mu^{1/2} r^2 
\end{eqnarray}
for $r\geq 0$. Similarly, for $s\geq 0$,
\begin{equation}
\lim_{L\to\infty} \frac 1{L^3} \big[  e(\mu+s) -e(\mu) + s\, 
\Tr\, P_\mu\big]  \geq - \frac 1{8\pi^2} \mu^{1/2} s^2
\left(1+\frac s\mu\right)^{1/2}\,.
\end{equation} 
Now if we choose $r= 4\pi^2 \mu^{-1/2}L^{-3} \Tr\, P_\mu(1-\gamma)$
and $s=0$, (\ref{442}) implies that 
\begin{equation}
\Tr\, (-\Delta-\mu)\gamma \geq e(\mu) + 2\pi^2 \mu^{-1/2} 
L^{-3} \left[\Tr\, P_\mu(1-\gamma)\right]^2  -o(L^3) \\  \label{ref1}
\end{equation}
for any $0\leq \gamma\leq 1$. 
On the other hand, choosing $r=0$ and  
\begin{equation}
s= \frac {4\pi^2}{L^3}  \frac { \Tr\, (1-P_\mu)\gamma}{\mu^{1/2}  
+ 2\pi^{-1/3}L^{-1}  \left( \Tr\, (1-P_\mu)\gamma \right)^{1/3}}\,,  
\end{equation}
a simple estimate yields 
\begin{equation}\label{ref1a}
\Tr\, (-\Delta-\mu)\gamma \geq e(\mu)   + \frac{\pi^2}{L^3}  
\frac{\left[\Tr\, (1-P_\mu)\gamma\right]^2}{\mu^{1/2} + 2\pi^{-1/3}L^{-1}  
\left( \Tr\, (1-P_\mu)\gamma \right)^{1/3}} -o(L^3)\,.
\end{equation}

Using (\ref{onepf}), we have 
\begin{equation}\label{ref2}
\Tr\, (-\Delta-\mu)\gamma- \frac 1\beta \SsFF[\gamma] \geq 
- \frac {L^3}4 P_0^L(\beta/2,\mu) + \frac 12 \Tr\, (-\Delta-\mu)\gamma\,.
\end{equation}
We use this estimate, together with (\ref{ref1}), on the last term on the right side of
(\ref{apr1}), with $\gamma=\gamma_\upa^{X^\doa}$.  We restrict our
attention again to states with $\Pp^L[\Gamma]\geq P_0^L(\beta,\mu)- C
a\rhov_0^2$ as above. Note that $e(\mu)=-\half L^3 P_0^L(\infty,\mu)$.
Let $\Delta P_0^L(\beta,\mu)$ denote the expression
\begin{equation}
\Delta P_0^L(\beta,\mu)\equiv - 2 P_0^L(\beta,\mu) + 
P_0^L(\beta/2,\mu) + P_0^L(\infty,\mu) \geq 0 \,.
\end{equation}
The positivity follows from convexity of
$P_0^L(\beta,\mu)$ in $1/\beta$. From (\ref{apr1}),
(\ref{ref2}) and (\ref{ref1}) we obtain
\begin{equation}\label{450}
\frac 1{L^6} \intsum dX^\doa\, n(X^\doa)\, \left[\Tr\, P_\mu
\big(1-\gamma_\upa^{X^\doa}\big)\right]^2  \leq \frac{\mu^{1/2}}{4\pi^2} 
\left( 4 C a\rhov_0^2 + \Delta P_0^L(\beta,\mu)\right) + o(1)\,.
\end{equation}
By repeating this argument, this time with (\ref{ref1a}) in place of
(\ref{ref1}), and using convexity of the map $x\mapsto
x^2/(1+x^{1/3})$, we also obtain the bound
\begin{multline}\label{451}
\frac { \left[ \frac 1{L^3} \intsum dX^\doa\, n(X^\doa)\, \Tr\, 
(1- P_\mu)\gamma_\upa^{X^\doa}\right]^2}{1+2\pi^{-1/3}\mu^{-1/2}
\left[ \frac 1{L^3} \intsum dX^\doa\, n(X^\doa)\, \Tr\, (1- P_\mu)
\gamma_\upa^{X^\doa}\right]^{1/3}} \\ \leq\frac{\mu^{1/2}}{2\pi^2} 
\left( 4 C a\rhov_0^2 + \Delta P_0^L(\beta,\mu)\right) + o(1)\,.
\end{multline}

We claim that 
\begin{equation}\label{labdel}
\Delta P_0^L(\beta,\mu) \leq \frac 2{3\pi^2} \frac{\mu^{1/2}}{\beta^2} 
\left( 1+\frac 1{\beta\mu}\right) + o(1) 
\end{equation}
in the thermodynamic limit. To see this, first note that $\Delta
P_0^L(\beta,\mu)\leq P_0^L(\beta/2,\mu)-P_0^L(\infty,\mu)$. Moreover,
\begin{eqnarray}\nonumber
\lim_{L\to\infty}\big[ P_0^L(\beta,\mu)-P_0^L(\infty,\mu)\big] \!\!\!
&=&\!\!\! \frac 1{(2\pi)^3\beta} \int_{p^2\leq \mu} dp\, \ln 
\left( 1+ z^{-1} e^{\beta p^2} \right) \\ \nonumber &&\!\!\! + \frac 1{(2\pi)^3\beta} 
\int_{p^2\geq \mu} dp\, \ln \left( 1+ z e^{-\beta p^2} \right)\,. \\ \label{97}
\end{eqnarray}
Estimating $\ln (1+x)\leq x$, we see that the first integral is
bounded by
\begin{equation}
\frac {4\pi}3 \frac {\mu^{1/2}}{z} \int_0^{\mu^{1/2}} dp\, p e^{\beta p^2}
 = \frac {2\pi}3 \frac{\mu^{1/2}}{\beta}\frac {z-1}{z} \leq 
\frac {2\pi}3 \frac{\mu^{1/2}}{\beta}\,.
\end{equation}
In a similar way, the second integral is bounded by
\begin{equation}\label{99}
\frac {4\pi}3 \frac z {\mu^{1/2}} \int_{\mu^{1/2}}^\infty dp\, 
p^3 e^{-\beta p^2} = \frac {2\pi}3 \frac{\mu^{1/2}}{\beta}
\left(1+ \frac{1}{\beta\mu}\right)\,.
\end{equation}
Hence we arrive at (\ref{labdel}). 

Now assume, as explained above, that $\beta\rhov_0^{2/3} >
(a^3\rhov_0)^{-1/9}$. For small $a^3\rhov_0$, this means that $z$ has
to be large. In this case,
\begin{equation}\label{456}
\frac{\mu^{1/2}}{2\pi^2} \left( 4 C a\rhov_0^2 + 
\Delta P_0^L(\beta,\mu)\right) \leq \const 
\rhov_0^2 \big(a\rhov_0^{1/3}\big)^{2/3}\,.
\end{equation}
Note that, for $P$ a projection,
\begin{eqnarray}\nonumber 
\!\!\!\!\!\!\! \|\gamma-P\|_1 &\leq&  \|(\gamma-1)P\|_1+\|\gamma(1-P)\|_1  
\\  \nonumber &\leq& \|P\|_2  \|(\gamma-1)P\|_2 + \|\gamma^{1/2}\|_2 
\|\gamma(1-P)\|_2 \\ &\leq& \|P\|_2   \big[ \Tr\, (1-\gamma)P\big]^{1/2} 
+ \|\gamma^{1/2}\|_2 \big[ \Tr\, \gamma(1-P)\big]^{1/2} \,. 
\end{eqnarray}
Moreover,
\begin{equation}\label{458}
\|\gamma^{1/2}\|_2^2 = \Tr\, \gamma \leq \Tr\, P + \Tr\, \gamma(1-P)\, .
\end{equation}
Note that $\Tr\, P_\mu \leq L^3 (6\pi^2)^{-1} \mu^{3/2}$ (using
Lemma~\ref{lemtrin}). By combining the estimates (\ref{450}),
(\ref{451}) and (\ref{456})--(\ref{458}), with $P=P_\mu$ and
$\gamma=\gamma_\upa^{X^\doa}$, we obtain that, for small $a^3\rhov_0$
and $\beta\rhov_0^{2/3} > (a^3\rhov_0)^{-1/9}$,
\begin{equation}\label{suppl}
\frac 1{L^3} \intsum dX^\doa\, n(X^\doa)\, \left\| 
P_\mu-\gamma_\upa^{X^\doa}\right\|_1 \leq \const \big(a\rhov_0^{1/3}\big)^{1/6}\,. 
\end{equation}
This inequality supplements (\ref{tres3}) in the case of large $z$.

It remains to estimate $\|P_\mu - \gamma_0\|_1= \Tr\,
P_\mu(1-\gamma_0) + \Tr\, \gamma_0 (1-P_\mu)$. In the thermodynamic
limit,
\begin{eqnarray}\nonumber
\lim_{L\to\infty} \frac 1{L^3} \|P_\mu - \gamma_0\|_1 &=& 
\frac 1{(2\pi)^3} \int dp\, \left[ \frac {\theta(\mu-p^2)}
{1+z e^{-\beta p^2}} +  \frac {\theta(p^2-\mu)}{1+z^{-1} e^{\beta p^2}}\right] 
\\ \nonumber 
&\leq & \frac 1{(2\pi)^3} \int dp\, \left[ \frac {\theta(\mu-p^2)}
{z e^{-\beta p^2}} +  \frac {\theta(p^2-\mu)}{z^{-1} e^{\beta p^2}}\right]
\\ &\leq & \frac 1{6\pi^2} \frac{\mu^{1/2}}\beta \left( 1+\frac 1{2\beta\mu}\right)\,,
\end{eqnarray}
where the last inequality is derived in the same way as in
(\ref{97})--(\ref{99}). In case $\beta\rhov_0^{2/3} >
(a^3\rhov_0)^{-1/9}$, as considered here, this last expression is
actually bounded by $(a\rhov_0^{1/3})^{1/3}$, and therefore negligible
compared with the error term on the right side of (\ref{suppl}) for
small $a^3\rhov_0$.

\subsubsection{Summary of this subsection}\label{sss4}

To summarize, we have shown in this subsection that for a density
matrix $\Gamma$ satisfying $\Pp^L[\Gamma]\geq P_0^L(\beta,\mu)- C
a\rhov_0^2$ we have the {\it a priori} bound
\begin{equation}\label{summar} 
\intsum dX^\doa\, n(X^\doa)\, \left\|
  \gamma_\per-\gamma_\upa^{X^\doa}\right\|_1 \leq C_\nu L^3 \rhov_0 
\big(a\rhov_0^{1/3}\big)^{1/6-\vu} + o(L^3)
\end{equation}
for $\vu>0$, for some constant $C_\nu$ depending only on $v$ (and $C$
above), but not on $z$.  Here, $\gamma_{\rm per}$ denotes the
one-particle density matrix of a system of non-interacting fermions
(at inverse temperature $\beta$ and chemical potential $\mu$) with
periodic boundary conditions on the cube $\Lambda_L$.

As a side note, we remark that (\ref{summar}) implies, in particular,
that the reduced one-particle density matrix of the dilute interacting
system, $\gamma_\upa$, is close to the one for non-interacting
particles. This, in turn, proves the inequality (\ref{thm2}) in
Corollary~\ref{C1}, with a worse error term than the one given in
(\ref{thm2}), however.

\subsection{Putting Things Together}\label{ssl4}

We now show how the estimates of the preceding subsections can be
combined to prove the desired lower bound on the pressure, given in
Theorem~\ref{T1}.  Let $h^\chi$ be the one-particle operator given in
(\ref{defhchi}).  It follows from (\ref{finsub}), (\ref{ineq1})
and (\ref{ineq2}) that, for any density matrix $\Gamma$,
\begin{eqnarray}\nonumber 
\!\!\!\! - \half L^3 \Pp^L[\Gamma] \!\!\!&\geq& \!\!\!  \delta\,
  \Tr_{\Hh_1}[-\Delta \gamo_\upa] + \, \Tr_{\Hh_1} \left[\left(
      h^\chi- \mu\right)\gamo_\upa\right] - \frac 1\beta
  \SsFF[\gamo_\upa] \\ && \!\!\! + (1-\delta)(1-\kappa) \intsum dX^\doa\,
  n(X^\doa)\, \Tr_{\Hh_1}\left[ W_{X^\doa} \gamo_\upa^{X^\doa}\right]\,.
\label{things} 
\end{eqnarray} 
Here, we used that $\Gamma$ is symmetric with respect to exchange of
the $\upa$ and $\doa$-particles, by assumption, which implies in
particular that $\gamma_\doa=\gamma_\upa$.  With the notation
introduced in Lemma~\ref{oplem},
\begin{equation}
 \Tr_{\Hh_1} \left[\left(
      h^\chi- \mu\right)\gamo_\upa\right] - \frac 1\beta
  \SsFF[\gamo_\upa] \geq \frac 1\beta f\big( \beta(h^\chi-\mu)\big)\,.
\end{equation}

In the last term in (\ref{things}), we use that
\begin{equation}
\Tr_{\Hh_1}\left[ W_{X^\doa}
\gamo_\upa^{X^\doa}\right] \geq \Tr_{\Hh_1}\left[ W_{X^\doa}
\gamo_\per\right] - \|W_{X^\doa}\|_\infty \|\gamo_\upa^{X^\doa}
-\gamo_\per\|_1\,.
\end{equation}
We have already estimated $\|W_{X^\doa}\|_\infty$ in (\ref{normW}),
and this estimate is independent of $X^\doa$. Moreover, the
$X^\doa$-average of $\|\gamo_\upa^{X^\doa}-\gamo_\per\|_1$ has been
estimated in (\ref{summar}). Because of translation invariance,
$\gamma_\per$ has a strictly constant density, given by $L^{-3} \Tr\,
\gamo_\per$. Thus
\begin{equation}\label{lastint}
\Tr_{\Hh_1}\left[ W_{X^\doa}
\gamo_\per\right] = \frac {\Tr\, \gamo_\per}{L^3} 
\int_{[0,L]^3} dx \, W_{X^\doa}(x)\,.
\end{equation}
We note that $L^{-3} \Tr\, \gamo_\per = \half \rhov_0 + o(1)$ in the
thermodynamic limit. To estimate the integral in (\ref{lastint}), we
write $W_{X^\doa}=W_+-W_-$, where $W_+$ denotes the positive terms in
(\ref{defwy}) containing $U$, and $-W_-$ the negative ones containing
$w_R$. For the negative part, we can use (\ref{intw}) to get the upper
bound
\begin{equation}
\int_{[0,L]^3} dx \, W_-(x) \leq \const \frac{a R^2}{\eps s^2} 
 |X^\doa|\,.
\end{equation}
Here, $|X^\doa|=M$ denotes the number of spin down particles.  Note
that
\begin{equation}
 \intsum dX^\doa\,
  n(X^\doa)\, |X^\doa| = \Tr\, \gamma_\doa\,.
\end{equation}
This trace can be estimated using (\ref{summar}), which in particular implies that 
\begin{equation}
\left|L^{-3}\Tr_{\Hh_1}\, \gamo_\doa - \half
\rhov_0\right| \leq C_\nu L^3 \rhov_0 
\big(a\rhov_0^{1/3}\big)^{1/6-\vu} + o(L^3)\, . 
\end{equation}
Moreover, using $\int U = 4\pi$,
\begin{eqnarray}\nonumber
\int_{[0,L]^3} dx \, W_+(x) &=& \sum_{\{ k\, : \, x^\doa_k \in 
\widetilde X^\doa_R\}}  (1-\eps) a \int_{[0,L]^3} dx\, U(x-x_k^\doa) 
\\ &\geq&   (1-\eps)4\pi a  \left[ \big|\widetilde X^\doa_R \big| - 
\const \frac {L^2}{R^2} \right]
\end{eqnarray}
for any fixed $X^\doa$. Here, $|\widetilde X^\doa_R |$ denotes the
number of elements in $\widetilde X^\doa_R$, i.e., the number of
$x_k^\doa$'s in $X^\doa$ whose distance to the nearest neighbor among
the $x_l^\doa$'s for $l\neq k$ is bigger than $2R$.  The last term in
square brackets bounds the number of $x_k^\doa$'s that are closer than
a distance $R$ to the boundary of the box. Since the distance between
the $x^\doa_k$'s in $\widetilde X^\doa_R$ is bigger than $2R$, the
number of such $x^\doa_k$'s close to the boundary is bounded by
$\const L^2/R^2$.

We now need an estimate on $ |\widetilde X^\doa_R |$. Note that 
\begin{equation}
 \big|\widetilde X^\doa_R \big| \geq  \big|X^\doa\big| - 
(2R)^2 \sum_{k=1}^M \frac 1{\delta_k^2}\,,
\end{equation}
where $\delta_k$ denotes the distance of $x_k^\doa$ to its nearest
neighbor in $X^\doa$. The last expression can be bounded from below
using the operator inequality
\begin{equation}
\sum_{k=1}^{M} \frac 1{\delta_k^2} \leq c
\sum_{k=1}^{M} -\Delta_k^\doa\,,
\end{equation}
which holds on the anti-symmetric tensor product $\bigwedge^{M}
L^2(\R^3)$, and is proved in \cite[Thm.~5]{LYau}. Here, $c$ is some
positive constant satisfying $c\leq 48$.  We thus have that
\begin{equation}
\intsum dX^\doa\, n(X^\doa)\,  \big|\widetilde X^\doa_R \big| \geq \Tr_{\Hh_1} 
\left(1 + c (2R)^2 \Delta\right) \gamo_\doa\,. 
\end{equation}
The negative last term involving the kinetic energy can be canceled by
an appropriate choice of $\delta$ in (\ref{things}).

By choosing $\delta= 2\pi a c \rhov_0 (2R)^2$ and taking the
thermodynamic limit $L\to\infty$ in (\ref{things}), the above
estimates imply that
\begin{eqnarray}\nonumber
P(\beta,\mu) \!\!\!\! &\leq&\!\!\!\!  -\limsup_{L\to\infty} 
\frac 2{\beta L^3} f\big(  \beta (h^\chi-
\mu)\big)   - 2\pi a \rhov_0^2 \left(1-\delta-\eps-\kappa\right) 
\\ \nonumber && \!\!\!\!\!\!\!  + C_\nu a \rhov_0^2 
\left[ \frac{R^2}{\eps s^2} + \left( 1+ \frac 1{\rhov_0} 
\max\left\{ \frac{1}{R^3-R_0^3},\, \frac 1{\eps Rs^2}\right\} \right)
\big (a\rhov_0^{1/3}\big)^{1/6-\vu}\right]. \\ 
\label{things2}
\end{eqnarray}

It remains to estimate the first term on the right side of this
expression.  To this end, let $\Upsilon(p)= (1-\delta) p^2
(1-(1-\kappa)\chi(p)^2)$ (compare with (\ref{defhchi})). We claim that
\begin{equation}
\liminf_{L\to\infty} \frac 1{\beta L^3} f\big(  \beta (h^\chi-
\mu)\big) \geq \frac {-1}{(2\pi)^3 \beta} \int_{\R^3} dp\, \ln 
\big(1+ z \exp\big(-\beta \Upsilon(p)\big)\big)\,.  \label{claa}
\end{equation}
This can be seen using coherent states \cite{anal} as follows. Let $u$
be a real function supported in a cube of side length $\ell$, with
$\int dx\, u(x)^2 =1$, and let $\xi_{y,k}(x)=u(x-y)e^{ikx}$. Denoting
by $|\xi_{y,k}\rangle\langle \xi_{y,k}|$ the projection onto
$\xi_{y,k}$, we then have the resolution of identity $(2\pi)^{-3} \int
dk\, dy\, |\xi_{y,k}\rangle\langle \xi_{y,k}| = 1$. Since $x\mapsto
-x\ln x$ is concave for $0\leq x\leq 1$, this implies that, for any
fermionic one-particle density matrix $\gamma$,
\begin{equation}\label{cohent}
\SsFF[\gamma] \leq - (2\pi)^{-3}\int dk\, dy\, \left[ \rho(y,k) 
\ln \rho(y,k)+ \big(1-\rho(y,k)\big)\ln \big(1-\rho(y,k)\big) \right]\,,
\end{equation}
where $\rho(y,k) \equiv \langle \xi_{y,k} |\gamma|\xi_{y,k}\rangle$.
Note that $0\leq \rho(y,k)\leq 1$, and that $\Tr\,
\gamma=(2\pi)^{-3}\int dk\, dy\, \rho(y,k)$.  With
$\widehat\gamma(k)=(2\pi)^{-3}\int dx\, dx'\, \gamma(x,x')
\exp(ik(x'-x))$ denoting the Fourier transform of $\gamma$, we can
write
\begin{equation}
 \Tr\, h^\chi \gamma = \int dp\, \Upsilon(p) \widehat\gamma(p)\,.
\end{equation}
Moreover, a simple calculation shows that
\begin{equation}\label{coh}
\int dk\, dy\, \Upsilon(k) \rho(y,k) = (2\pi)^{3} \int dk\, dq\, 
\Upsilon(k+q) \widehat\gamma(k) |\widehat u(q)|^2 \,.  
\end{equation}
Note that $\widehat \gamma(k)\geq 0$.  In the case considered here,
$\Upsilon$ has a bounded Hessian, i.e., $\partial_i\partial_j
\Upsilon(k) \leq 2 C$ as a matrix, for some constant $C$ (depending only
on the parameter $s$ used in the definition (\ref{defchi}) of $\chi$).  Therefore
$\Upsilon(k+q)\leq \Upsilon(k)+ q \nabla\Upsilon(k) + C q^2$.
Inserting this estimate into (\ref{coh}) and noting that $\int dq\, q
|\widehat u(q)|^2=0$ since $u$ is assumed to be real, we obtain
\begin{equation}\label{cohfin}
\Tr\, [ h^\chi \gamma] \geq (2\pi)^{-3} \int dk\, dy\, 
\rho(y,k)\left( \Upsilon(k) - C \int dx\, |\nabla u(x)|^2\right) \,.
\end{equation} 
Now choose $u$ such that $\int dx\, |\nabla u(x)|^2\leq C' \ell^{-2}$ for some constant $C'$ independent of $\ell$.
Then (\ref{cohent}) and (\ref{cohfin}) imply that
\begin{eqnarray}\nonumber
&&\Tr\, [ (h^\chi-\mu) \gamma] - \frac 1\beta \SsFF[\gamma] 
\\ \nonumber &&\geq (2\pi)^{-3} \int dk\, dy\, 
\left( \Upsilon(k) - \mu - C C' \ell^{-2}\right) \rho(y,k)  
\\ \nonumber && \quad + \frac 1{(2\pi)^3 \beta}  
\int dk\, dy\, \left[ \rho(y,k) \ln \rho(y,k)+ \big(1-\rho(y,k)\big)
\ln \big(1-\rho(y,k)\big) \right]\,. \\ \label{479}
\end{eqnarray}
Note that $\rho(y,k)\neq 0$ only if $y$ is an cube of side length
$L+2\ell$. If we minimize the integrand for each fixed $y$ (in this
cube) and fixed $k$, we see that the right side of (\ref{479}) is
bounded from below by
\begin{equation}
- (L+2\ell)^3 \frac 1{(2\pi)^3\beta} \int dk\, 
\ln\left( 1+ z e^{\beta C C' /\ell^2} e^{-\beta \Upsilon(k)} \right)\,.
\end{equation}
Dividing by $L^3$, letting $L\to\infty$ and then $\ell\to\infty$ we
arrive at (\ref{claa}).
 
It remains to estimate the integral on the right side of (\ref{claa}),
and compare it with $-\half P_0(\beta,\mu)$. We do this in two steps.
First, note that
\begin{eqnarray}\nonumber 
&&\frac {2}{(2\pi)^3 \beta} \int dp\, \ln 
\left(1+ z e^{-\beta \Upsilon(p)}\right) - (1-\delta) 
P_0\big(\beta(1-\delta),\mu/(1-\delta)\big) \\ 
&& = \frac {2}{(2\pi)^3 \beta} \int dp\, \ln 
\frac{ 1+ z e^{-\beta \Upsilon(p)}}{ 1+ z e^{-\beta (1-\delta)p^2}}\,. \label{481}
\label{st1}
\end{eqnarray}
By the definition (\ref{defchi}) of $\chi$,
$\Upsilon(p)=(1-\delta)p^2$ for $|p|\leq 1/s$. Hence the integrand in
(\ref{481}) is only non-zero for $|p|\geq 1/s$, in which case
$\Upsilon(p)\geq (1-\delta) \kappa p^2$. Hence (\ref{st1}) is bounded
from above by
\begin{eqnarray}\nonumber 
&&\frac {2}{(2\pi)^3 \beta} \int_{|p|\geq 1/s} dp\, \ln 
\frac{ 1+ z e^{-\beta (1-\delta)\kappa p^2}}
{ 1+ z e^{-\beta (1-\delta)p^2}} \leq \frac {2z}{(2\pi)^3 \beta}
 \int_{|p|\geq 1/s} dp\, e^{-\beta (1-\delta)\kappa p^2} \\ 
&&\leq \frac 1{\sqrt{2} \pi^2} \frac 1{\beta^{5/2}} 
\frac 1{(1-2\delta)\kappa} e^{-\beta(  (1/2-\delta)\kappa s^{-2} - \mu)}\,, 
\end{eqnarray}
where we estimated $|p|\leq (2\beta)^{-1/2} \exp(\half\beta p^2)$ in
the integrand in order to evaluate the last integral. Secondly, by
simple scaling,
\begin{equation}
(1-\delta) P_0\big(\beta(1-\delta),\mu/(1-\delta)\big) 
= (1-\delta)^{-3/2} P_0(\beta,\mu)\,. 
\end{equation}
We thus arrive at the following estimate:
\begin{eqnarray}\nonumber
P(\beta,\mu) \!\!\! &\leq&\!\!\!  (1-\delta)^{-3/2} P_0(\beta,\mu)  
 - 2\pi a \rhov_0^2 \left(1-\delta-\eps-\kappa\right) \\ \nonumber 
&& \!\!\!\!\!  + C_\nu a \rhov_0^2 \left[ \frac{R^2}{\eps s^2} 
+ \left( 1+  \frac 1{\rhov_0} \max\left\{ \frac{1}{R^3-R_0^3}, \,
\frac 1{\eps Rs^2}\right\}\right) \big (a\rhov_0^{1/3}\big)^{1/6-\vu}\right] 
\\ &&  \!\!\!\!\!+  \frac 1{\sqrt{2} \pi^2} \frac 1{\beta^{5/2}} 
\frac 1{(1-2\delta)\kappa} e^{-\beta(  (1/2-\delta)\kappa s^{-2} - \mu)} \,,
\label{things3}
\end{eqnarray}
with $\delta$ given as above by $\delta= 2\pi a c \rhov_0 (2R)^2$.  It
remains to choose $s$, $R$, $\eps$ and $\kappa$. We take
\begin{equation}\label{ch1}
R=\rhov_0^{-1/3} \big(a\rhov_0^{1/3}\big)^{1/22}\, , \ s=\rhov_0^{-1/3} 
\big(a\rhov_0^{1/3}\big)^{1/66}\, , \ \eps = \big(a\rhov_0^{1/3}\big)^{1/33} 
\end{equation}
and
\begin{equation}\label{ch2}
\kappa = \big(a\rhov_0^{1/3}\big)^{1/33-\nu}\,.
\end{equation}
Note that with this choice of parameters, $\delta = 8\pi c
(a^3\rhov_0)^{12/33}$. Hence 
\begin{equation}
\delta P_0(\beta,\mu) = \const  a\rho_0^2 \big(a^3\rho_0\big)^{1/33} 
\frac{P_0(\beta,\mu)}{\rho_0(\beta,\mu)^{5/3}},
\end{equation}
and the last fraction is uniformly bounded for $z$ bounded away from
zero (as already argued in the lower bound after (\ref{347})).
Moreover, since $\kappa s^{-2} = \rhov_0^{2/3}(a^3\rhov_0)^{-\nu/3}$,
we see by the same reasoning as in the lower bound in (\ref{secl3})
that the last term in (\ref{things3}) is actually exponentially small
in $a\rhov_0^{1/3}$, uniformly in $z$ for bounded $1/z$.  Inserting
(\ref{ch1}) and (\ref{ch2}) into (\ref{things3}) above, we thus obtain
\begin{equation}
P(\beta,\mu)\leq P_0(\beta,\mu) - 2\pi a \rhov_0^2 \left( 1 -
C_\nu(z) \big (a\rhov_0^{1/3}\big)^{1/33-\nu}\right)
\end{equation}
for any $\nu>0$, with $C_\nu(z)$ bounded for bounded $1/z$. This
finishes the proof of the upper bound in Theorem~\ref{T1}.

\section{Proof of Corollaries~\ref{C1} and~\ref{C2}}\label{CP}

In this final section, we show how to derive Corollaries~\ref{C1}
and~\ref{C2} from Theorem~\ref{T1}. We start with proving
Corollary~\ref{C1}. The essential ingredient is convexity of
$P(\beta,\mu)$ in $\mu$. It implies that
\begin{equation}\label{cor11}
\rhov_-(\beta,\mu)\leq \rhov_+(\beta,\mu) 
\leq \frac {P(\beta,\mu+\delta)-P(\beta,\mu)}\delta
\end{equation}
for any $\delta>0$. 
Using (\ref{thm1}) as well as the fact that $\rhov_0(\beta,\mu)$ is
monotone increasing in $\mu$, this
yields
\begin{eqnarray}\label{rhoes}
\rhov_+(\beta,\mu) &\leq&  \frac{ P_0(\beta,\mu+\delta)-P_0(\beta,\mu) }{\delta} \\ \nonumber
&& + \frac 1 {\delta} 
\rhov_0(\beta,\mu+\delta)^{5/3} C_\alpha\big(e^{\beta \mu}\big)  
\big(a\rhov_0(\beta,\mu+\delta)^{1/3}\big)^{1+ \alpha}\,.
\end{eqnarray}
Again by convexity, $P_0(\beta,\mu+\delta)-P_0(\beta,\mu)\leq \delta
\rhov_0(\beta,\mu+\delta)$. We choose
\begin{equation}\label{choosedelta}
\delta = 
\big (a \rhov_0(\beta,\mu)^{1/3}\big)^{(1+\alpha)/2}\, \max\left\{\mu , \, 1/\beta\right\} \,.
\end{equation}
With this choice of $\delta$, it is then not difficult to see that 
\begin{equation}
\frac{\rhov_0(\beta,\mu+\delta)}{\rhov_0(\beta,\mu)} \leq 
1 + \const \big( a\rhov_0^{1/3}\big)^{(1+\alpha)/2}
\end{equation}
for some constant independent of $\mu$ and $\beta$. 
Using this estimate, (\ref{rhoes}) implies
that
\begin{equation}\label{cor12}
\rhov_+(\beta,\mu) \leq \rhov_0(\beta,\mu) \left( 1+ 
\widehat C_\alpha(z) \big( a\rhov_0^{1/3}\big)^{(1+\alpha)/2}\right)
\end{equation}
for some $\widehat C_\alpha(z)$ that is uniformly bounded for bounded
$1/z$.

A lower bound on $\rhov_-(\beta,\mu)$ can be obtained similarly, using
that, for $\delta>0$,
\begin{equation}
\rhov_-(\beta,\mu) \geq \frac {P(\beta,\mu-\delta)-P(\beta,\mu)}\delta\,.
\end{equation}
Proceeding along the same lines as in (\ref{cor11})--(\ref{cor12})
above, this proves Corollary~\ref{C1}.

Next we prove Corollary~\ref{C2}. For $\rho>0$, let
$f(\beta,\rho)=\sup_\mu [ \mu\rho - P(\beta,\mu)]$ denote the free
energy per unit volume.  If the supremum is attained at some $\hat\mu$, then
convexity of $P(\beta,\mu)$ in $\mu$ implies that
\begin{equation}\label{supr}
\rhov_-(\beta,\hat\mu)\leq
\rho\leq \rhov_+(\beta,\hat\mu)\,.
\end{equation}
Let $f_0(\beta,\rho)=\sup_\mu [ \mu\rho- P_0(\beta,\mu)]$ denote the
free energy density for the ideal Fermi gas. The supremum is achieved
at $\mu_0$, determined by $\rhov_0(\beta,\mu_0)=\rho$. Hence we
immediately get the lower bound
\begin{equation}\label{lbf}
f(\beta,\rho)\geq \mu_0 \rho - P(\beta,\mu_0) 
\geq f_0(\beta,\rho) + 2 \pi a \rho^2 - 
C_\alpha\big(e^{\beta\mu_0}\big) a \rho^2 \big (a\rho^{1/3}\big)^\alpha\,,
\end{equation}
where we used (\ref{thm1}) to estimate $P(\beta,\mu_0)$ in terms of
$P_0(\beta,\mu_0)$.

To get an upper bound on the free energy, we first make an {\it a
  priori} estimate to ensure that $\hat \mu$ is close to $\mu_0$.
Suppose that $\hat\mu < \mu_0$. Then, by (\ref{supr}) and monotonicity
of $\rhov_+(\beta,\mu)$ in $\mu$, $\rho=\rhov_0(\beta,\mu_0)\leq
\rhov_+(\beta,\mu_0-\delta)$ for all $\delta \leq \mu_0 - \hat\mu$.
Using (\ref{thm2}), this implies that
\begin{equation}\label{inse}
\rhov_0(\beta,\mu_0)\leq \rhov_0(\beta,\mu_0-\delta) \left(1+ 
K_\alpha(\beta,\mu_0-\delta)\right)
\end{equation}
for all $\delta \leq \mu_0 - \hat\mu$, 
where we denoted 
\begin{equation}
K_\alpha(\beta,\mu) =  
\widehat C_\alpha(e^{\beta\mu}\big) 
\big( a\rhov_0(\beta,\mu)^{1/3}\big)^{(1+\alpha)/2}\,.
\end{equation}
Ineq. (\ref{inse}) then implies that 
\begin{equation}\label{hd}
\delta \leq \const  
\big (a \rho^{1/3}\big)^{(1+\alpha)/2}\, \max\left\{\mu_0 , \, 1/\beta\right\} 
\end{equation}
for some constant independent of $\rho$ and $\beta$ (compare with
(\ref{choosedelta})). Denoting the right side of (\ref{hd}) by
$\bar\delta$, we therefore see that $\hat\mu \geq \mu_0 - \bar\delta$.

The same method works in the case when $\hat\mu > \mu_0$. One uses
that $\rhov_0(\beta,\mu_0)\geq \rhov_-(\beta,\mu_0+\delta)$ for all
$\delta \leq \hat \mu - \mu_0$. Proceeding along the same lines, this
implies that $\hat\mu \leq \mu_0 + \bar\delta$. In particular,
\begin{equation}\label{the1}
f(\beta,\rho) = \sup_{|\mu-\mu_0| \leq \bar\delta } \big[ \mu \rho - P(\beta,\mu)\big]\,.
\end{equation}
Using (\ref{thm1}) and (\ref{thm2}), it is not difficult to see that
\begin{equation}\label{the2}
P(\beta,\mu) \geq P_0(\beta,\mu) - 2\pi a \rhov_0(\beta,\mu_0)^2 
\left( 1  + \widetilde C_\alpha\big ( e^{\beta\mu_0}\big)
\big (  a\rhov_0(\beta,\mu)^{1/3}\big)^{\alpha} \right) 
\end{equation}
if $|\mu-\mu_0|\leq \bar\delta$, for some $\widetilde C_\alpha(z)$
that is uniformly bounded for bounded $1/z$.  Inserting the bound
(\ref{the2}) into (\ref{the1}) proves the desired upper bound on
$f(\beta,\rho)$.
 
Note that $e^{\beta\mu_0}$ is bounded away from zero for bounded
$1/(\beta\rho^{2/3})$. Hence (\ref{the2}) and (\ref{lbf}) imply the
statement in Corollary~\ref{C2}.

\section*{Acknowledgments}

It is a pleasure to thank Elliott Lieb for numerous fruitful
discussions.

\end{document}